\documentclass[preprint,review,12pt]{elsarticle}

\usepackage{amssymb}
\usepackage{latexsym}
\usepackage{amsfonts}
\usepackage{graphicx}
\usepackage{float}
\usepackage{amsmath}
\usepackage{mathcomp}
\usepackage{epsfig}
\usepackage{hyperref}
\usepackage{makecell}

\journal{Journal of Computational Physics}

\begin{document}

\begin{frontmatter}
\author{Isha Dhiman}
\author{Arvind Kumar Gupta \corref{cor1}}
\ead{akgupta@iitrpr.ac.in}
\cortext[cor1]{Corresponding Author. Tel:+ 91 1881 242140; Fax:+ 91 1881 223395}
\address{Department of Mathematics, Indian Institute of Technology Ropar,
    Rupnagar-140001, India \vspace{-1.0cm}}

\title{Collective dynamics of an inhomogeneous two-channel exclusion process: Theory and Monte Carlo simulations}

\begin{abstract}
This work is devoted to the development of a novel theoretical approach, named \emph{hybrid approach}, to handle a localized bottleneck in a symmetrically coupled two-channel totally asymmetric simple exclusion process with Langmuir kinetics. The hybrid approach is combined with singular perturbation technique to get steady-state phase diagrams and density profiles. We have thoroughly examined the role played by the strength of bottleneck, binding constant and lane-changing rate in the system dynamics. The appearances of bottleneck-induced shock, a bottleneck phase and Meissner phase are explained. Further, the critical values of bottleneck rate are identified, which signify the changes in the topology of phase diagram. It is also found that increase in lane-changing rate as well as unequal attachment, detachment rates weaken the bottleneck effect. Our theoretical arguments are in good agreement with extensively performed
Monte Carlo simulations.
\end{abstract}
\end{frontmatter}

\section{Introduction}
\label{sec:1}
In protein synthesis, the genetic information is deciphered into proteins by molecular machines called ribosomes that attach themselves at the start end of mRNA, move along the chain in a unidirectional manner and finally detach at the stop end~\cite{chou2004clustered}. Each translation step requires the binding of a freely diffusing transfer-RNA (tRNA) molecule, carrying the amino acid specific to each codon~\cite{chou2004clustered,shaw2003totally}.  The important factor affecting the ribosome translation rate is relative concentrations of tRNA, which may vary from codon to codon. The codons with lower concentrations of tRNA reduce the protein synthesis rate and thus plays the role of an inhomogeneity in a homogeneous system~\cite{solomovici1997does,sorensen1989codon,stenstrom2001codon}. Apart from this, inhomogeneities also occur naturally in many other transport systems such as vehicular traffic~\cite{gupta2014phase}, blood flow~\cite{nichols2011mcdonald} and flow of data in a Von Neumann architecture~\cite{lin2013monte}. In traffic flow, the ongoing construction on roads, a slow moving vehicle or an accident can lead to slow down the flow rate on highways and can lead to congestion. Further, the separation of the CPU and the memory in computers creates Von Neumann bottleneck, which limits the performance of the computer via limited bandwidth between the CPU and the memory.

Totally asymmetric simple exclusion process (TASEP)~\cite{chowdhury2000statistical,macdonald1968kinetics} is well known to be a paradigmatic model for studying stochastic transport in many-particle systems. Both single-channel and multi-channel TASEPs have been well explored theoretically as well as using Monte Carlo simulations~\cite{derrida1998exactly,gupta2013coupling,krugboundary1991,popkov2001symmetry,pronina2004two,shi2011strong}. Further, one has to take into account the fact that the proteins as molecular motors can also attach from the bulk reservoir or detach from it (Langmuir Kinetics (LK))~\cite{howardsinauer}. In contrast to TASEP, where total number of particles remain conserved, the additional attachment-detachment dynamics (LK dynamics) violate the particle conservation and leads to many interesting phenomena~\cite{parmeggiani2004totally}. In literature, the consequences of coupling of the two different dynamics : TASEP and LK have been well analyzed in single-channel~\cite{mirin2003effect,mukherji2006bulk,parmeggiani2004totally} as well as two-channel homogeneous systems~\cite{dhiman2014effect,gupta2014asymmetric,jiang2007two,wang2007effects}.
The idea of slowing down of particles at certain defect positions can be incorporated in the form of a set of inhomogeneous lattice sites (bottleneck), either as a single unit or randomly distributed over the whole lattice in TASEP. This type of disorder is known as site-wise disorder. Another type of disorder studied in literature is particle-wise, where a slow moving particle itself acts as an inhomogeneity in the system.
The present work focuses on site-wise disorder which is suitable to model the inhomogeneities present in transport of molecular motors.

Although a lot of work has been done on homogeneous TASEPs, the effect of disorder on the steady-state dynamics of such systems is not well understood. Several studies have been performed on single-channel inhomogeneous TASEPs with~\cite{pierobon2006bottleneck,qiu2007density} as well as without LK~\cite{chou2004clustered,dong2007towards,greulich2008phase,janowsky1994exact,janowsky1992finite,kolomeisky1998asymmetric,shaw2003totally,tripathy1998driven}. While investigating the role of a bottleneck in a closed TASEP~\cite{janowsky1994exact,janowsky1992finite}, it was found that even the presence of a single bottleneck site can produce shock profile and a plateau in the fundamental current-density relation. Later, Kolomeisky~\cite{kolomeisky1998asymmetric} examined even richer case of single-channel open system and explored the consequences of the inhomogeneity with both faster and slower transition rates. He divided the system into two homogeneous TASEPs coupled at the single bottleneck site (defect mean-field theory (DMFT)) and proved analytically that a fast site has no effect on the phase diagram; whereas a slow site leads to shifting of the phase boundaries only. He also tested the theoretical results with Monte Carlo simulations and found a good agreement in low density (LD) and high density (HD) phases; while a little deviation in maximal current (MC) phase. Another analytical approach namely finite-segment mean-field theory (FSMFT) was introduced by Chou and Laktaos~\cite{chou2004clustered} to study clusters of slow codons in protein synthesis. They found that ribosome density profiles near neighboring clusters of slow codons
suppress the proteins synthesis. Dong et al.~\cite{dong2007towards} generalized the DMFT to study the effect of two bottlenecks on the protein production rate and also performed extensive Monte Carlo simulations to conclude that the location as well as spacing between the bottlenecks affect the production rate, pointing out an important phenomenon namely \textit{edge effect}. The investigation into edge effects was further carried out by Greulich and Scadshneider~\cite{greulich2008phase} to generate the phase diagrams of inhomogeneous TASEP using interacting subsystem approximation (ISA). They could successfully explain the interactions of defects with the boundaries of the single-channel system.
The more complex case of inhomogeneous TASEP in the presence of Langmuir kinetics was studied by Qiu et al.~\cite{qiu2007density}. They calculated phase diagrams and density profiles by adopting the concept of DMFT and also studied the effect of slow hopping rate and detachment rate on the phase diagram. Pierobon et al.~\cite{pierobon2006bottleneck} provided a detailed study on the role of a bottleneck in a TASEP with LK, using an effective mean-field theory and Monte Carlo simulations. They introduced the concept of carrying capacity to identify various novel phases called bottleneck phases. Importantly, all of the above studies focused on single-channel inhomogeneous systems.

Parallel to the inhomogeneous single-channel systems, particles in multi-channel transport systems~\cite{dhiman2014effect,gupta2013coupling,gupta2014asymmetric,jiang2008weak,pronina2004two,shi2011strong} may also confront a bottleneck, present in either one or in more than one channels. The importance of studying the multi-channel system lies in the fact that it can act as a framework for extending the analysis to networks. Due to the complexity in dynamics generating from the interactions between different channels, it is difficult to examine the effects of inhomogeneity in a multi-channel open system. Up to our knowledge, the only contribution in this direction has been made by Wang et al.~\cite{wang2008local}, which explored the effect of a local inhomogeneity in one of the lanes of a two-lane TASEP with LK under a symmetric lane changing rule. They extended the DMFT to a two-channel system by incorporating the concept of effective injection and removal rates at the inhomogeneous lattice site. Despite a good agreement between the solution of the mean-field equations and Monte Carlo simulations in ref.~\cite{wang2008local}, this approach fails to produce the steady-state phase diagrams of the two-channel system. Moreover, it is not feasible to analyze the role played by various parameters such as lane-changing rate, attachment-detachment rate and strength of bottleneck on the steady-state dynamics. One can infer from here that though DMFT is capable to provide analytic solutions for single-channel inhomogeneous TASEPs with~\cite{pierobon2006bottleneck,qiu2007density} as well as without LK~\cite{kolomeisky1998asymmetric}, it lacks some important ingredients to generate a complete picture of the dynamics of the corresponding two-channel system as discussed in section ~\ref{sec:3}. This motivates us to develop a new approach to handle the bottleneck in a two-channel TASEP with LK, which not only overcomes the existing limitations, but also demonstrates the unexplored dynamics of two-channel inhomogeneous systems.

The objective of the proposed study is two-fold, to develop a general theoretical approach, which is capable to produce the phase diagrams and to study the effect of various system parameters on the steady-state phases. In this paper, we attempt to provide a complete picture of the dynamics of two-channel symmetrically coupled TASEP with LK in the presence of a single localized bottleneck in one of the two channels by adopting a new and simplified approach, called the \emph{hybrid approach}. We have also validated the theoretical results with Monte Carlo simulations. The paper is organized as follows. In section~\ref{sec:2}, we define the model under examination and the governing dynamical rules. We briefly discuss the limitations of the earlier approaches in sec.~\ref{sec:3}. The theoretical hybrid approach and Monte Carlo simulations are covered in section~\ref{sec:4} and section~\ref{sec:5}. A thorough analysis of the stationary properties of the model is discussed in section~\ref{sec:6}. In the concluding section~\ref{sec:7}, we summarize the results and future prospectives of our work.
\section{Two-channel inhomogeneous TASEP with LK}
\label{sec:2}
\begin{figure}
\centering
\includegraphics[width=11.5cm,height=4.5cm]{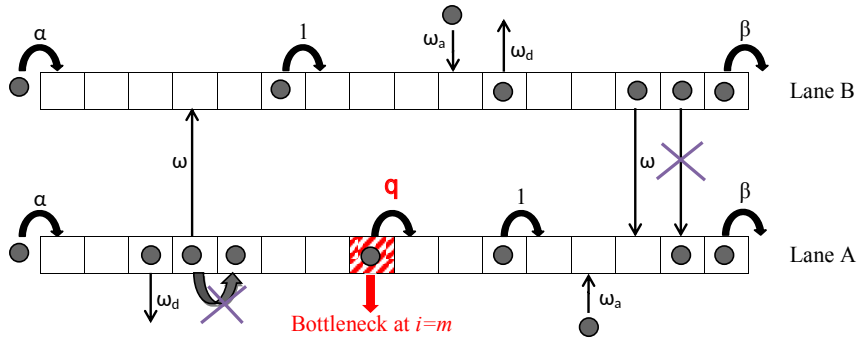}
\caption{Schematic diagram of a symmetrically coupled two-channel TASEP with LK with bottleneck in lane A at $m^{th}$ site. Crossed arrows show forbidden transitions.}\label{fig:1}
\end{figure}

We define our model in a two-channel $(L,2)$ lattice, where $L$ is the length of a channel. The two channels are denoted by $A$ and $B$, in which particles are distributed under hard-core exclusion principle
(See Fig.~\ref{fig:1}). We adopt random-sequential update rules for the dynamical evolution of the system. For each time step, a lattice site $(i,j)$; $i=1,2,3,.....L$; $j= A,B$
is randomly chosen. The state of the system is characterized by a set of occupation numbers $
\tau_{i,j}$ ($i=1,2,3,.....L$; $j= A,B$), each of which is either zero (vacant site) or one
(occupied site). At entrance ($i=1$), a particle can enter the lattice with a rate $\alpha$
provided $\tau_{1,j}=0$; and at exit ($i=L$), a particle can leave the lattice with a rate $\beta$
when $\tau_{L,j}=1$. In the bulk, if $\tau_{i,j}=1$, then the particle at the site $(i,j)$ firstly
tries to detach itself from the system with a rate $\omega_d$ (detachment rate) and if it fails then it
moves forward to site $(i+1,j)$ with a rate $p_{i,j}$ provided $\tau_{i+1,j}=0$; otherwise it attempts to shift to other
lane with a rate $\omega$, only if the target
site is vacant. On the other hand, if $\tau_{i,j}=0$ ; $i= 2,3,....,L-1$, a particle can attach to the site $(i,j)$ with a rate $\omega_a$ (attachment rate). Here, horizontal transition rate $p_{i,j}$ is inhomogeneous and is given by the following binary distribution
\begin{equation}
p_{i,j}=\left\{\begin{array}{ll} q~; & ~~~~i=m~~ \& ~~j=A\\ 1~; &~~~~\text{otherwise} \end{array}\right.\label{eq:1}
\end{equation}
It is clear from Eq.~\eqref{eq:1} that the bottleneck connects the sites at $i=m$ and $i=m+1$ in lane $A$ (Fig.~\ref{fig:1}). Here, $q$ denotes the transition rate of a particle on passing through the bottleneck and will be called as \textit{bottleneck rate} throughout this paper. We wish to analyze the effects of a localized bottleneck in the bulk with no boundary interactions of the open system. This adds up to the assumption viz., $1 << m << L$.
\section{Inadequacies of DMFT}
\label{sec:3}
The only step to handle a single-site bottleneck in a two-channel TASEP with LK was put forth by Wang et al.~\cite{wang2008local}. They generalized the concept of DMFT from a single-channel inhomogeneous TASEP with LK to the corresponding two-channel system. Although inhomogeneity is present in only one lane, yet they divided both the lanes into two sub-lanes separated at the position of bottleneck site and introduced effective exit and entrance rates at bottleneck site, respectively in each lane. Using mean-field approximation with some simplified assumptions, they computed the nonlinear system of equations in terms of effective rates and solved it numerically. The steady-state density profiles obtained using DMFT are found to be in good agreement with Monte Carlo simulations. But the incompetency of this approach lies in the fact that the nonlinear system may not have a unique feasible solution always for all choices of parameters, due to which the complete information about the steady-state dynamics is missing. Moreover, the phase diagrams cannot be obtained and it is not feasible to quantify the role played by different parameters such as lane changing rate, attachment-detachment rates and bottleneck rate etc. Since this approach uses the concept of dividing each lane (inhomogeneous as well as homogeneous) into subsystems, its generalization either to multi-channel system or to include multiple defect sites would be a cumbersome job to handle.
\section{Theoretical approach}
\label{sec:4}
As discussed in the previous section that DMFT is incapable of explaining certain important aspects, we introduce a novel \textit{hybrid approach} to deal with the bottleneck in a multi-channel TASEP. Instead of dividing both the lanes into two sublanes each, we divide only the inhomogeneous lane (lane A) into two subsystems viz., $1 \leq i < m$ and $ m+1 < i \leq L$, connected through sites $i=m$ and $i=m+1$.

Firstly, we compute the temporal evolution of occupation probabilities $\tau_{i,j}$ in the bulk ($1<i<L$, $j=A,B$) from the following set of master equations:
\begin{equation}
\begin{split}
\frac{d\langle \tau_{i,j}\rangle}{dt} = p_{i-1,j}\langle \tau_{i-1,j}(1-\tau_{i,j})\rangle- p_{i,j}\langle \tau_{i,j}(1-\tau_{i+1,j})\rangle + \omega_a \langle 1-\tau_{i,j}\rangle -\omega_d\langle \tau_{i,j}\rangle \\
\mp \omega \bigg(\langle \tau_{i,A}\tau_{i+1,A}(1-\tau_{i,B})-\langle \tau_{i,B}\tau_{i+1,B}(1-\tau_{i,A})\rangle\bigg),\label{eq:2}
\end{split}
\end{equation}
where $\langle\cdots\rangle$ denotes the statistical average
and last term on right-hand side takes a negative and a positive sign for lane $A$ and $B$, respectively.
At the boundaries, the particle densities evolve according to
\begin{eqnarray}
\frac{d\langle \tau_{1,j}\rangle}{dt}=\alpha\langle 1-\tau_{1,j}\rangle - p_{1,j}\langle \tau_{1,j}(1-\tau_{2,j})\rangle, \label{eq:3}\\
\frac{d\langle \tau_{L,j}\rangle}{dt}= p_{L,j}\langle \tau_{L-1,j}(1-\tau_{L,j})\rangle - \beta\langle \tau_{L,j}\rangle. \label{eq:4}
\end{eqnarray}
Factorizing the correlations using mean-field approximation, we get
\begin{math}
\langle \tau_{i,j}\tau_{i+1,j}\rangle=\langle \tau_{i,j}\rangle \langle \tau_{i+1,j}\rangle \end{math}.
\subsection{Continuum mean-field equations}
To find the continuum limit of the master equations, we define lattice constant $\epsilon=1/L$, rescale the time  and other kinetic rates as $t{'}=t/L$, $\Omega_a=\omega_a L, \Omega_d=\omega_d L$ and $\Omega=\omega L$~\cite{gupta2014asymmetric,parmeggiani2004totally}.

Writing $x=(i-1)\epsilon$ , we replace binary discrete variables $\tau_{i,j}$ with continuous variables $\rho_{i,j}\in [0,1]$ and retain the terms up to second-order in Taylor's series expansion to obtain
\begin{equation}
\rho_{i\pm 1,j}=\rho_{i,j}\pm \epsilon\frac{\partial\rho_{i,j}}{\partial x}+ \frac{\epsilon^2}{2}\frac{\partial^2\rho_{i,j}}{\partial x^2}+ O(\epsilon^3).\label{eq:7}
\end{equation}

\subsubsection{Continuum mean-field equation for lane B} Since lane B is free from any inhomogeneity, we can drop the subscript $i$ and compute the following time evolution equation from master equation~\eqref{eq:2} for average density in lane B, denoted by $\rho_B$.
\begin{equation}
\begin{split}
\frac{\partial \rho_B}{\partial t{'}}=&\frac{\epsilon}{2}\frac{\partial ^2 \rho_B}{\partial x^2}+\frac{\partial}{\partial x}(\rho_B^2-\rho_B)+ \Omega_d (K-(K+1)\rho_B)+ \Omega \big(\rho_A^2(1-\rho_B)\\&- \rho_B^2(1-\rho_A)\big),\label{eq:8}
\end{split}
\end{equation}
where $K=\Omega_a/\Omega_d$ is the binding constant and an important parameter in analysing the system dynamics~\cite{parmeggiani2004totally}.
\subsubsection{Hybrid mean-field system of equations for lane A} We divide lane A into two sublattices viz., $1 < i < m$ and $ m+1 < i < L$, connected through sites $i=m$ and $i=m+1$. The two homogeneous subsystems connected by the bottleneck in lane A will be referred to as left and right subsystem. Since, each of the two subsystems is individually a homogeneous TASEP with LK, we can find the continuum limit of the mean-field approximate equation for each subsystem in lane A by proceeding in a similar fashion as for lane B. Keeping the master equations at $(m,A)$ and $(m+1,A)$ sites intact after rescaling of aforementioned variables, we obtain the following \emph{hybrid} system of equations for time evolution of average density in lane A, denoted by $\rho_A$.

\textbf{Hybrid system}\\
\textit{Continuum part:} For $x \in (0, m\epsilon) \bigcup ((m+1)\epsilon, 1)$, we have
\footnotesize
\begin{equation}
\begin{split}
\frac{\partial \rho_A}{\partial t{'}}&=\frac{\epsilon}{2}\frac{\partial ^2 \rho_A}{\partial x^2}+\frac{\partial}{\partial x}(\rho_A^2-\rho_A)+ \Omega_d (K-(K+1)\rho_A)- \Omega \big(\rho_A^2(1-\rho_B)\\&- \rho_B^2(1-\rho_A)\big).\label{eq:9}
\end{split}
\end{equation}
\normalsize
\textit{Discrete part:}
\footnotesize
\begin{eqnarray}
\begin{split}
\frac{\partial\rho_{m,A}}{\partial t{'}}&=\frac{1}{\epsilon} \rho_{m-1,A}(1-\rho_{m,A})-\frac{q}{\epsilon} \rho_{m,A}(1-\rho_{m+1,A})+ \Omega_d (K-(K+1)\rho_{m,A})\\&- \Omega\big(\rho_{m,A}\rho_{m+1,A}(1-\rho_{m,B})-\rho_{m,B}\rho_{m+1,B}(1-\rho_{m,A})\big),\\
\frac{\partial\rho_{m+1,A}}{\partial t{'}}&= \frac{q}{\epsilon} \rho_{m,A}(1-\rho_{m+1,A})- \frac{1}{\epsilon} \rho_{m+1,A}(1-\rho_{m+2,A})+ \Omega_d (K-(K+1)\rho_{m+1,A})\\&- \Omega\big(\rho_{m+1,A}\rho_{m+2,A}(1-\rho_{m+1,B})-\rho_{m+1,B}\rho_{m+2,B}(1-\rho_{m+1,A})\big).\label{eq:10}
\end{split}
\end{eqnarray}
\normalsize
In the continuum limit, the boundary equations \eqref{eq:3} and \eqref{eq:4} reduce to $\rho_A(0)=\rho_B(0)=\alpha$ and $\rho_A(1)=\rho_B(1)=1-\beta$. Along with these boundary conditions, the system of equations \eqref{eq:8}, \eqref{eq:9} and \eqref{eq:10} is consistent for which, a unique feasible solution always exists. With this favour, our new approach overcomes the limitation of DMFT, which was unable to guarantee a unique feasible solution to the nonlinear system of equations~\cite{wang2008local}. Moreover, no approximation, other than mean-field, has been used to generate the hybrid system, which is again an advantage over the methodology of DMFT~\cite{wang2008local} for a two-channel inhomogeneous TASEP with LK. It can be easily seen that the hybrid approach only segments the inhomogeneous lane into two parts, without disturbing the homogeneous lane. This idea makes it applicable to more general systems such as multi-channel or networks and those with multiple bottleneck sites.
\subsection{Steady-state solution}
We introduce a new space variable $\widehat{x}$ to merge the equations~\eqref{eq:8},\eqref{eq:9} and \eqref{eq:10} into a single system, which further reduces to the following system in steady-state.
\footnotesize
\begin{eqnarray}
\begin{split}
\frac{\epsilon}{2}\frac{d^2 \rho_A}{d \widehat{x}^2}+(2\rho_A-1)\frac{d\rho_A}{d\widehat{x}}+\Omega_d (K-(K+1)\rho_A)-\Omega \rho^{2}_A (1-\rho_B)+\Omega \rho^{2}_B(1-\rho_A)&=&0,\\
\rho_{m-1,A}(1-\rho_{m,A})-q \rho_{m,A}(1-\rho_{m+1,A})+\omega_d (K-(K+1)\rho_{m,A})\\-\omega \rho_{m,A}\rho_{m+1,A}(1-\rho_{m,B})+\omega \rho_{m,B}\rho_{m+1,B}(1-\rho_{m,A}) &=&0,\\
q \rho_{m,A}(1-\rho_{m+1,A})-q \rho_{m+1,A}(1-\rho_{m+2,A})+\omega_d (K-(K+1)\rho_{m+1,A})\\-\omega \rho_{m+1,A}\rho_{m+2,A}(1-\rho_{m+1,B})+\omega \rho_{m+1,B}\rho_{m+2,B}(1-\rho_{m+1,A})&=&0,\\
\frac{\epsilon}{2}\frac{d^2 \rho_B}{dx^2}+(2\rho_B-1)\frac{d\rho_B}{dx}+\Omega_d (K-(K+1)\rho_B)+\Omega \rho^{2}_A (1-\rho_B)-\Omega \rho^{2}_B(1-\rho_A)&=&0, \label{eq:11}
\end{split}
\end{eqnarray}
\normalsize
where $0 < \widehat{x} < m\epsilon$, $ (m+1)\epsilon < \widehat{x} < 1$ and $0 < x < 1$.

Since it is difficult to solve the system ~\eqref{eq:11} analytically due to its hybrid nature, we propose to find the steady-state solution of the hybrid system using the singular perturbation technique~\cite{cole1968perturbation}.
Recently, this technique has been successfully applied for studying two-channel homogeneous TASEP with LK~\cite{dhiman2014effect,gupta2014asymmetric}.

Singular perturbation technique typically involves obtaining solutions of the differential equations describing the boundary layer region and the bulk solution, separately and then matching both the solutions to get the global solution. The bulk part of the solution, known as outer solution, is found in the limit $\epsilon\rightarrow 0$. We propose a generalized methodology to find the outer solution of the hybrid system~\eqref{eq:11}. Instead of solving the system~\eqref{eq:11} explicitly, we find the steady-state outer solution by capturing the long time solution of \eqref{eq:8}, \eqref{eq:9} and \eqref{eq:10} using the scheme discussed below~\cite{dhiman2014effect,gupta2014asymmetric}. We discretize the continuum part of model equations using finite-difference and replace the time derivative in discrete equations at sites $(m,A)$ and $(m+1,A)$ with forward-difference formula. For $1<i<m$, $m+1<i<L$ with $j=A$ and $1<i<L$ with $j=B$, the following scheme is applied to model equations~\eqref{eq:8}, \eqref{eq:9} and \eqref{eq:10} with the boundary conditions $\rho_A(0)=\rho_B(0)=\alpha$ and $\rho_A(1)=\rho_B(1)=1-\beta$.
\footnotesize
\begin{eqnarray}
\begin{split}
\rho_{i,j}^{n+1} = & \rho_{i,j}^n+ \frac{\epsilon}{2}\frac{\Delta t{'}}{\Delta x^2}\big({\rho_{i+1,j}^n-2\rho_{i,j}^n+\rho_{i-1,j}^n}\big) +
\frac{\Delta t{'}}{2\Delta x}\big[(2\rho_{i,j}^n-1)\big(\rho_{i+1,j}^n-\rho_{i-1,j}^n\big)\big]\\&+
\Delta t{'}\big[\Omega_d (K-(K+1)\rho_{i,j}^n) \mp\Omega \big((\rho_{i,A}^n)^2(1-\rho_{i,B}^n)- (\rho_{i,B}^n)^2(1-\rho_{i,A}^n)\big)\big],\\
\rho_{m,A}^{n+1} = & \rho_{m,A}^n+ \Delta t{'}\big[\frac{q}{\epsilon}\rho_{m-1,A}^n(1-\rho_{m,A}^n)-\frac{1}{\epsilon} \rho_{m,A}^n(1-\rho_{m+1,A}^n)+\Omega_d (K-(K+1)\\&\rho_{m,A}^n)-\Omega \rho_{m,A}^n \rho_{m+1,A}^n(1-\rho_{m,B}^n)+\Omega \rho_{m,B}^n \rho_{m+1,B}^n(1-\rho_{m,A}^n) \big],\\
\rho_{m+1,A}^{n+1} = & \rho_{m+1,A}^n+ \Delta t{'}\big[\frac{q}{\epsilon}\rho_{m,A}^n(1-\rho_{m+1,A}^n)- \frac{1}{\epsilon} \rho_{m+1,A}^n(1-\rho_{m+2,A}^n)+\Omega_d (K-(K+1)\\&\rho_{m+1,A}^n) -\Omega \rho_{m+1,A}^n \rho_{m+2,A}^n(1-\rho_{m+1,B}^n)+\Omega \rho_{m+1,B}^n \rho_{m+2,B}^n(1-\rho_{m+1,A}^n) \big].\label{eq:12}
\end{split}
\end{eqnarray}
\normalsize
The solution is captured in the limit $n \rightarrow \infty$ to ensure the occurrence of a steady-state. Clearly, the above system is consistent and gives a unique steady-state outer solution. It is easy to check the consistency of the above system of equations by analyzing the system in the limit $q \rightarrow 1$.

To satisfy the boundary conditions at both the ends, the density profiles incurs a crossover narrow regime in the form of either a boundary layer or a shock. This solution is known as
inner solution and is found by rescaling the space variable in the neighbourhood of $x_d$ as $\widetilde{x}=\frac{x-x_d}{\epsilon}$, where $x_d$ is the position of boundary layer. This rescaling eliminates the non-conservative source terms, which are formed by lane-changing transitions and attachment-detachment dynamics, in the system~\eqref{eq:11}. In terms of $\widetilde{x}$, the inner solution $\rho_{j,in}$ is given by
\begin{equation}
\frac{d \rho_{j,in}}{d \widetilde{x}}=2(a_j+\rho_{j,in}-\rho_{j,in}^{2}).\label{eq:13}
\end{equation}
Here, the integration constant $a_j$ is computed from the matching condition of outer and inner solution.
For example, let the boundary layer appears at right boundary ($x=1$) in lane $j$, the matching condition requires
\begin{math}
\rho_{j,in}(\widetilde x \rightarrow -\infty)=\rho_{j,out}(x=1)=\rho_{j,o}.
\end{math}
Here, $\rho_{j,o}$ is value of the outer solution in lane-$j$ at $x=1$. Clearly, $\rho_{j,o}$
is a function of system parameters $\Omega_d$, $\Omega_a$, and $\Omega_A$.
Solving Eq.~\eqref{eq:13} with $a_j=\rho_{j,o}^{2}-\rho_{j,o}$; we get
\begin{equation}
\rho_{j,in}=\frac{1}{2}+\frac{|2\rho_{j,o}-1|}{2}\tanh\bigg(\widetilde x |2\rho_{j,o}-1|+\xi_j\bigg), \label{eq:14}
\end{equation}
where $\xi_j=\tanh^{-1}\bigg(\frac{1-2\beta}{|2\rho_{j,o}-1|}\bigg)$.
The solution given by eq.~\eqref{eq:14} represents a right boundary layer (rbl) in lane $j$ with positive slope ($\tanh-r$). When $\beta <\rho_{j,o}$, the inner solution fails to satisfy the right boundary condition
$\rho_{j,in}(\widetilde x\rightarrow \infty)=1-\beta$ and deconfines from the boundary to enter the bulk of
lane-${j}$ in the form of a shock. Thus $\beta=\rho_{j,o}(\alpha)$ acts as a bulk phase transition line. Apart from deconfinement, the inner solution also undergoes change in its slope across the line $\beta=1-\rho_{j,o}(\alpha)$. The right boundary layer with negative slope is given by \begin{math}
\rho_{j,in}=\frac{1}{2}+\frac{|2\rho_{j,o}-1|}{2}\coth\bigg(\widetilde x |2\rho_{j,o}-1|+\hat{\xi}_j\bigg), \label{eq:15}
\end{math}
where $\hat{\xi}_j= \coth^{-1}\bigg(\frac{1-2\beta}{|2\rho_{j,o}-1|}\bigg)$.
Here, the change in the slope of boundary layer describes a surface transition, which does not affect
bulk density profile. Analyzing, in a similar fashion, we can find mathematically all the bulk as well as surface transition lines in the $\alpha-\beta$ plane and generate the phase diagram for different parameters. For a more detailed solution methodology of singular perturbation technique, we refer~\cite{dhiman2014effect}.
\section{Monte Carlo simulations}
\label{sec:5}
To testify the outcomes of our theoretical mean-field approach, we simulate the two-channel model using Monte Carlo simulations (MCS)~\cite{Bortz197510}. We adopt random sequential update dynamical rules, which is the realization of the usual master equation in continuous time. Moreover, random-sequential type update rule is appropriate to model the dynamics of intracellular protein transport~\cite{parmeggiani2004totally,dhiman2014effect}. We run the Monte Carlo simulations for $10^{10} - 10^{11}$ time steps and first $5\%$ steps are ignored
to ensure the occurrence of steady-state. The densities
in both the lanes have been computed by taking time
averages over an interval of $10 L$. Since the size of a real system is normally not very large, it is reasonable to simulate the system for a lattice size up to $L=1000$. Note that the dynamical rules involve many kinds of particle hoppings such as horizontal hoppings, vertical transitions and attachment-detachment dynamics, it is computationally expensive to calculate steady-state densities. As we proceed in the next section, it is found that the results of theoretical mean-field approach agree to a good extent with Monte Carlo simulations.
\section{Results and discussion}
\label{sec:6}
In this section, we provide a detailed discussion of the steady-state properties of the proposed model, with emphasis on the resulting phase diagrams and analyze the effects of bottleneck strength, lane-changing rate and binding constant. For simplicity, we assume that the bottleneck in lane A is located exactly at the middle i.e. $m=L/2$. The steady-state phase diagrams as well as density profiles are obtained from hybrid mean-field approach and validated using extensive Monte Carlo simulations.

When there is no bottleneck, i.e. $q=1$, the present model reduces to the homogeneous two-channel TASEP with LK. Under symmetric coupling conditions, the two-channel homogeneous TASEP with LK is analogous to its single-channel counterpart, which has been studied well in literature~\cite{wang2007effects}. For $q>1$, we call the defect to be \emph{fast defect} as the bottleneck rate is more than the usual horizontal particle hopping rate in the system. Clearly, a single fast defect does not affect the steady-state dynamics as the rate-limiting hoppings occurring with rate $1$ dominate the whole lattice~\cite{kolomeisky1998asymmetric}. Thus, it is appropriate to focus on the situation where the bottleneck acts as a slow defect i.e. $q < 1$.

It is well known that the steady-state dynamics of exclusion process are significantly affected by the binding constant. So, we proceed by segmenting the analysis into two subcases viz.~(i) $K=1$, and (ii) $K \neq 1$. The other parameters are kept fixed to $\Omega_d=0.2$ and $L=1000$.
\subsection{$K=1$}
\begin{figure}[htb]
\centering
   \includegraphics[width=6.75cm,height=5.8cm]{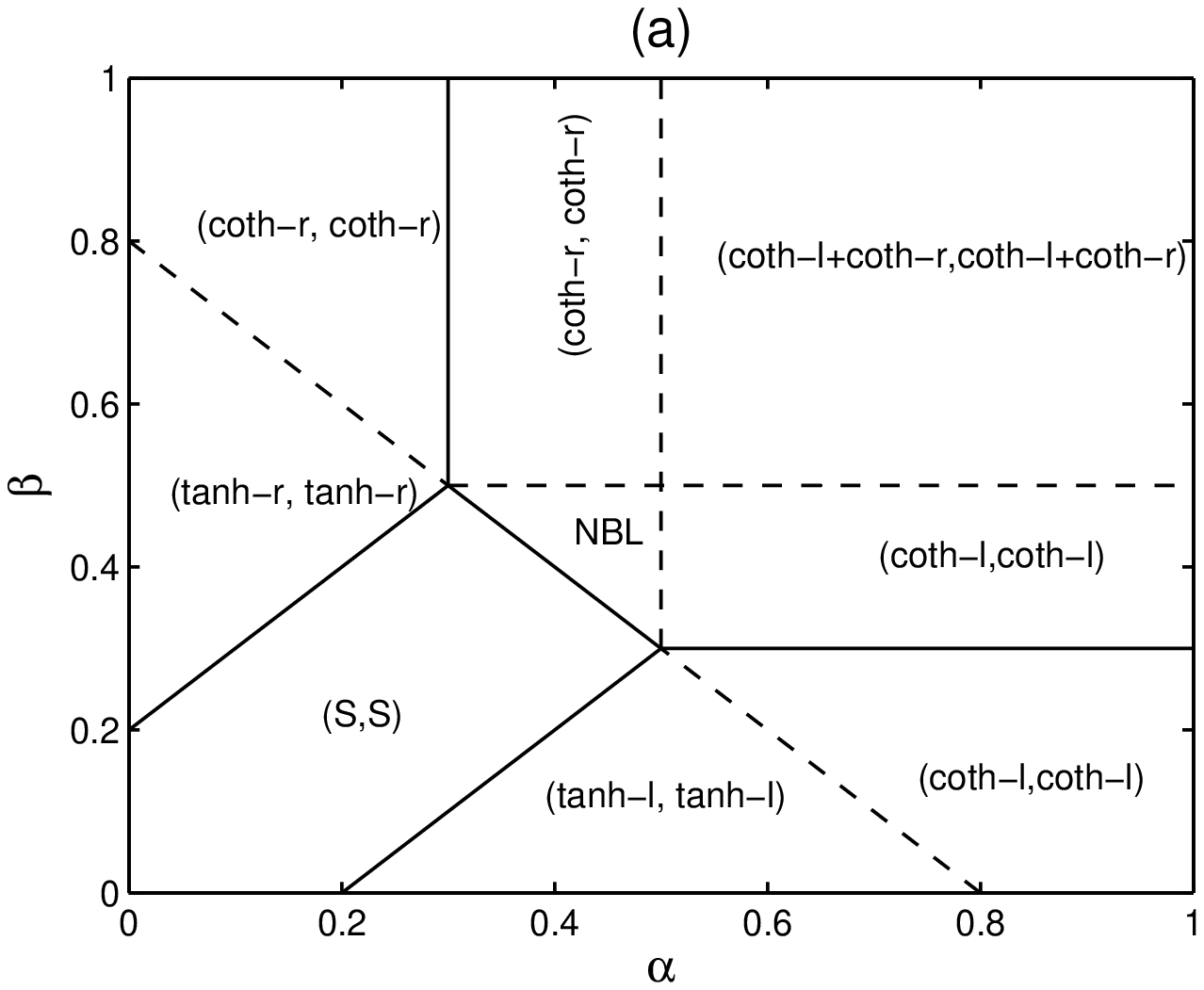}
    \includegraphics[width=6.75cm,height=5.8cm]{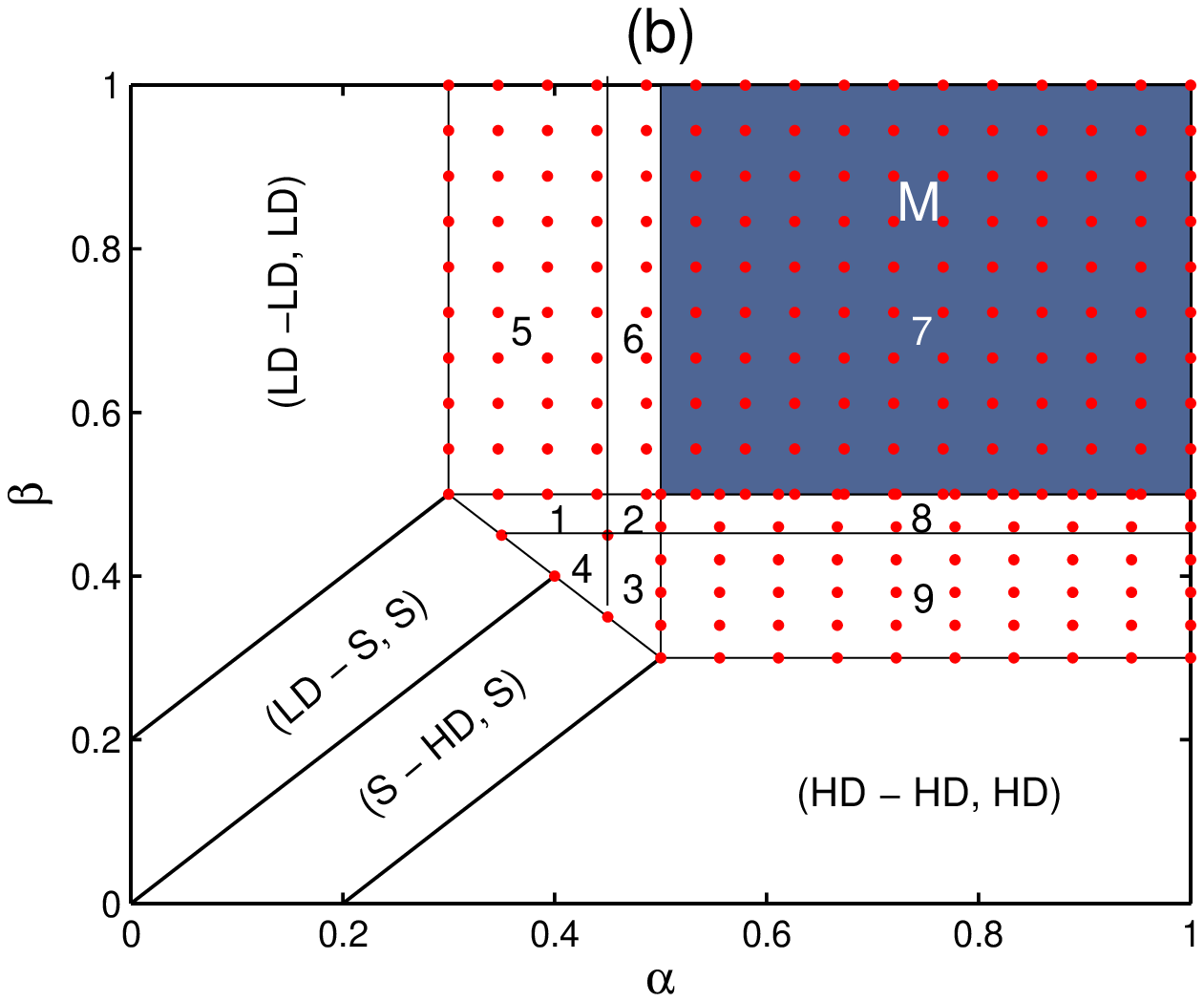}
  \caption{For $q=0.75$, $\Omega_d=0.2$, $\Omega=1$ and $L=1000$, (a) Phase diagram
showing bulk as well as surface transitions and classification according to the boundary layers, NBL denotes no boundary layer, (b) Phase diagram showing qualitative nature of phases with no surface transitions. Notations:- M: Meissner phase, 1: ($S_b$-LD/MC/HD,LD/HD/LD/MC/HD), 2: (LD/MC/HD-LD/MC/HD,LD/MC/HD/LD/MC/HD), 3: (LD/MC/HD-S,LD/MC/HD/LD/HD), 4: ($S_b$-S,LD/HD/LD/HD), 5: ($S_b$-LD/MC,LD/HD/LD/MC), 6: (LD/MC/HD-LD/MC,LD/MC/HD/LD/MC), 7:(MC/HD-LD/MC,MC/HD/LD/MC), 8: (MC/HD-LD/MC/HD,MC/HD/LD/MC/HD) and 9: (MC/HD-S,MC/HD/LD/HD). Solid and dashed lines denote bulk and surface transitions, respectively.}\label{fig:2}
\end{figure}

We examine the steady-state dynamics for different strengths of bottleneck by gradually decreasing the value of $q$. No significant qualitative changes are found in the phase diagram till $q_{c,1}\approx 0.75$, below which one notices the effect of bottleneck more clearly. Fig.~\ref{fig:2} shows the phase diagram of the system with $q=0.75$ and $\Omega=1$. Specifically, fig.~\ref{fig:2}(a) shows the complete characterization of various phases in terms of the boundary layers. The solid and dashed lines denote the bulk and surface transitions, respectively. As an illustration, the notation (tanh-r,tanh-r) signifies that there is a right boundary layer in lane A as well as lane B of the kind tanh-r, given by eq.~\eqref{eq:14} and $S$ denotes the shock phase. For a detailed understanding of the bulk and surface transitions, we refer the readers to references~\cite{dhiman2014effect,mukherji2006bulk}. As discussed in previous section that a surface transition is responsible for the change in slope of the boundary layer only and does not affect the bulk density, henceforth, our focus will be only on bulk transitions to avoid any complexities in the phase diagrams. Fig.~\ref{fig:2}(b) represents the bulk phase transitions only corresponding to fig.~\ref{fig:2}(a).

To understand the dynamics in a better way, we introduce an appropriate notation $(Ph_{A(left)}-Ph_{A(right)}, Ph_B)$ denoting the kind of stationary phase in the two-channel system. Here, $Ph_{A(left)}(Ph_{A(right)})$ denotes the type of stationary phase in left (right) subsystem of lane A; while $Ph_{B}$ denotes the stationary phase in lane B. For example, the phase (LD-S,S) denotes that the region left to the bottleneck is low density (LD) with a shock present in the right subsystem in lane A and a shock is also present in lane B.
\begin{figure}
  \centering
   \includegraphics[trim=12 00 25 00,clip,height=4.75cm,width=5.25cm]{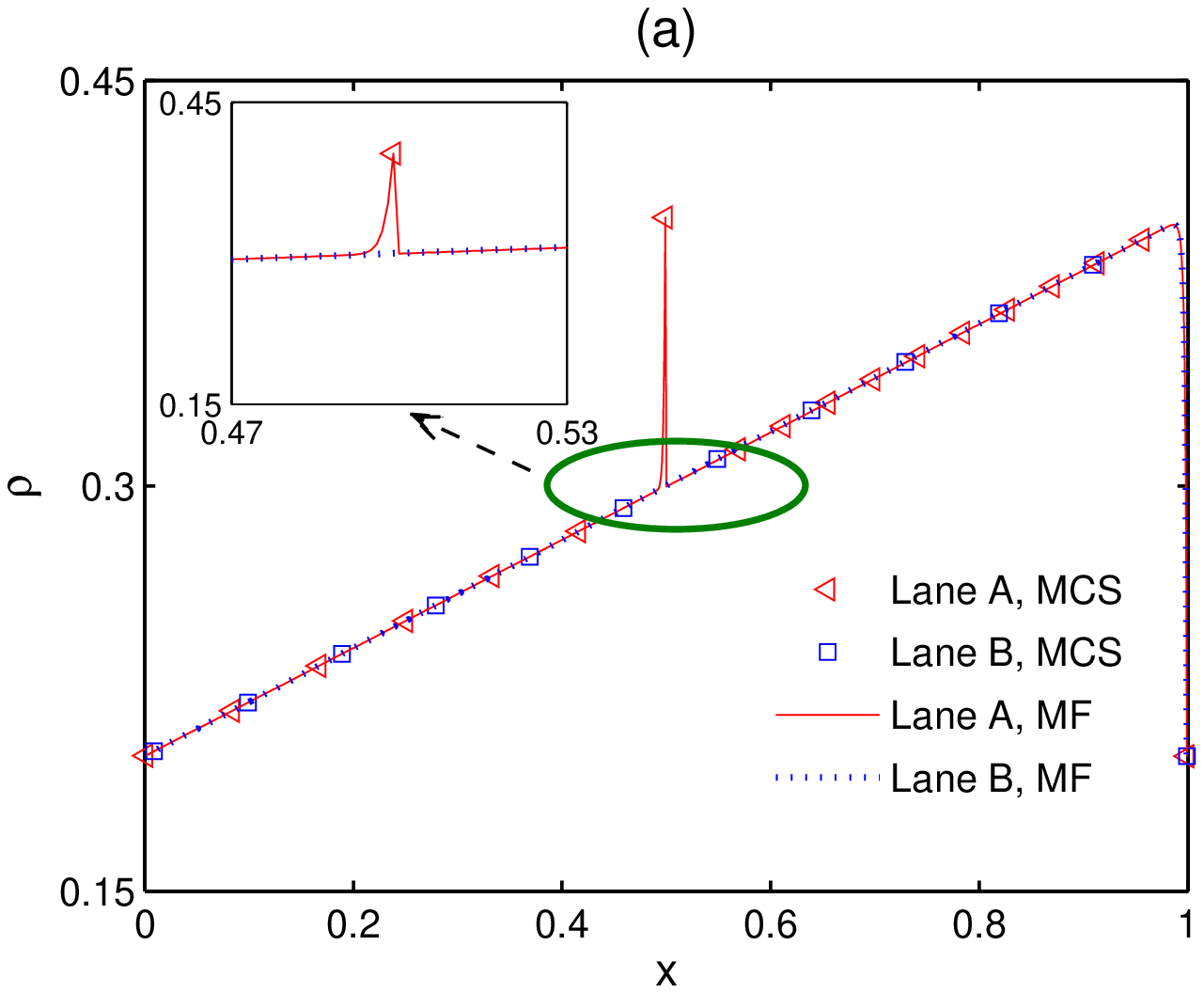}
   \includegraphics[trim=12 00 25 00,clip,height=4.75cm,width=5.25cm]{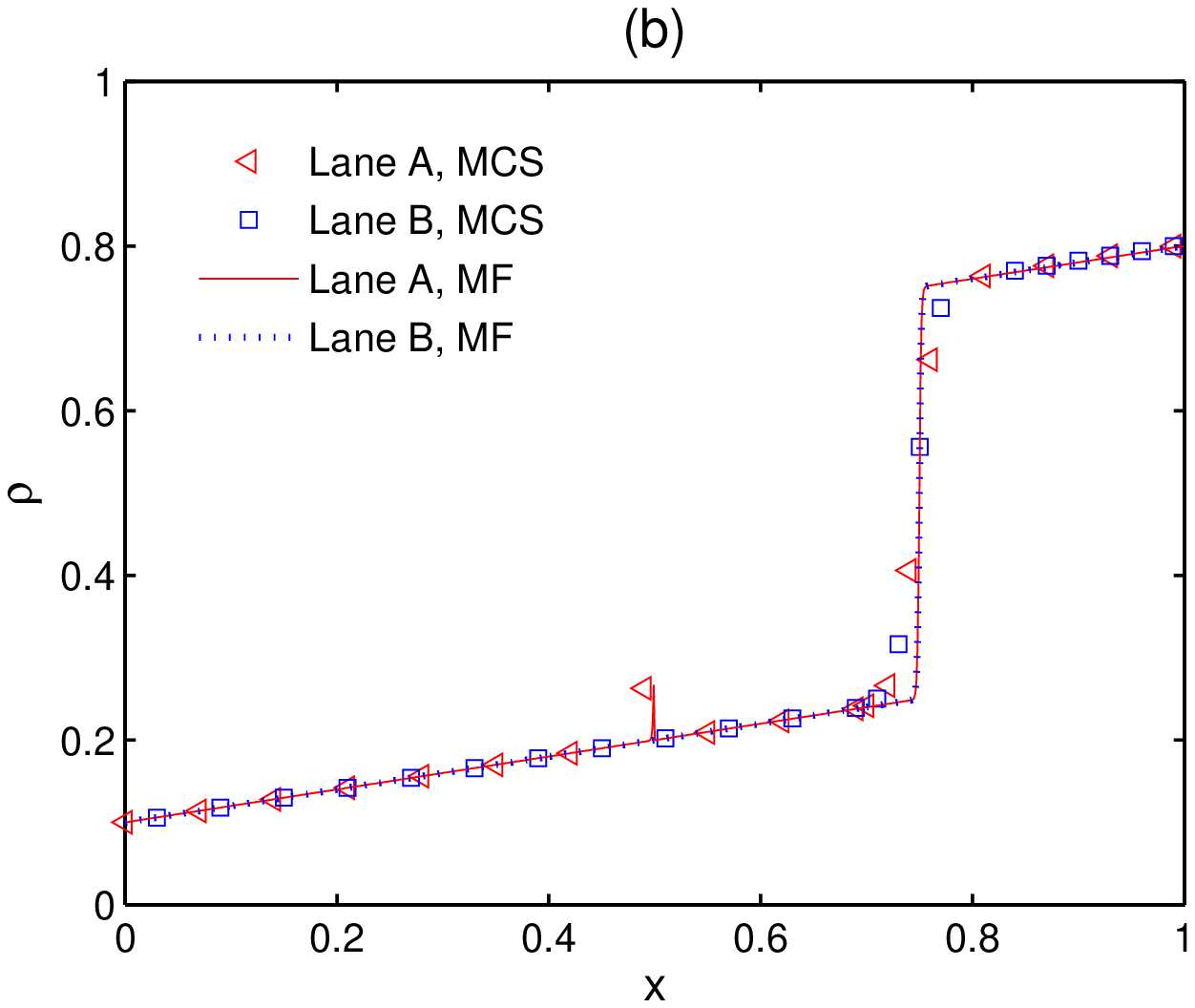}\\
   \includegraphics[trim=15 00 20 00,clip,height=4.75cm,width=5.25cm]{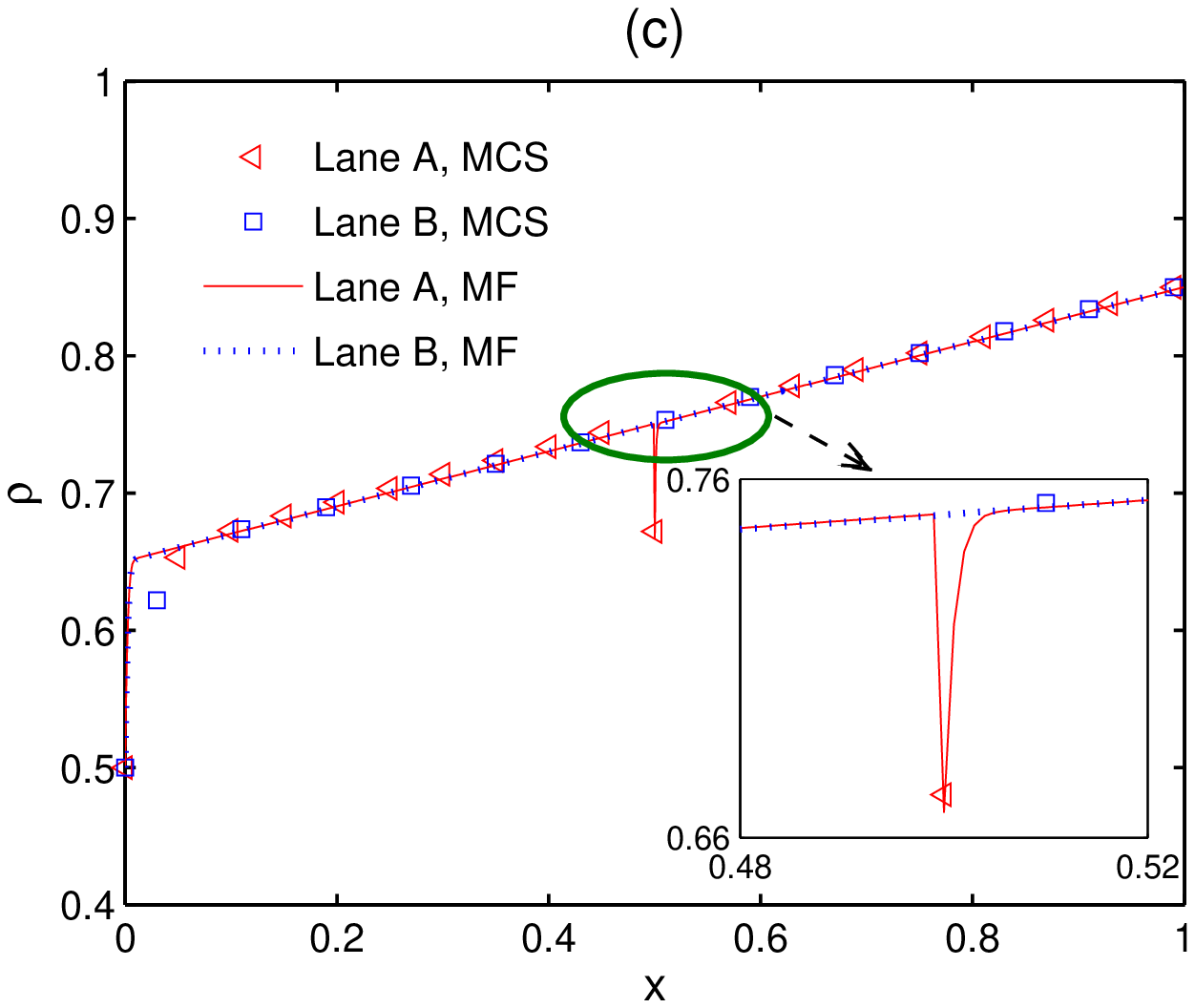}
   \includegraphics[trim=15 00 20 00,clip,height=4.75cm,width=5.25cm]{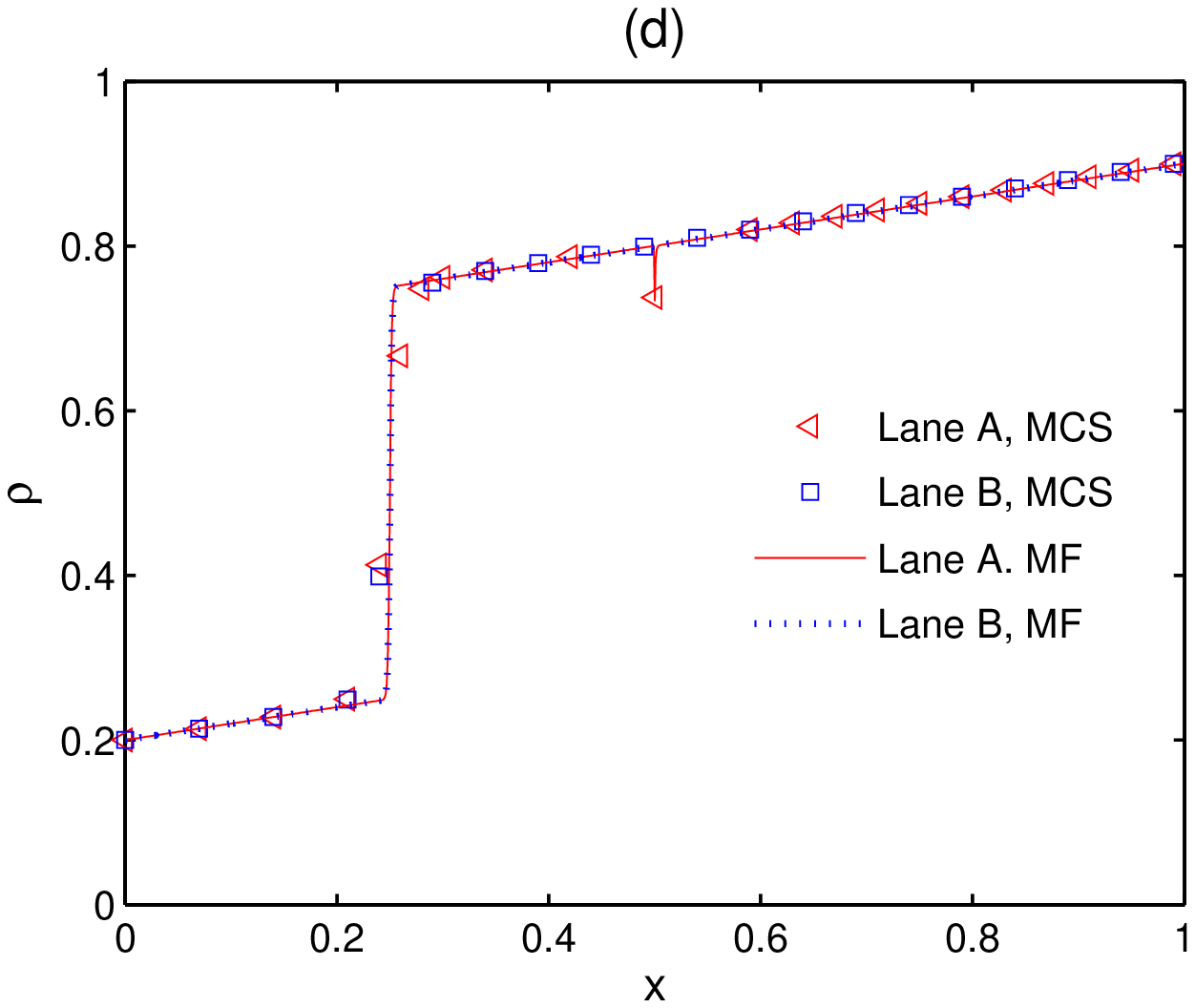}\\
   \includegraphics[trim=15 00 20 00,clip,height=4.75cm,width=5.25cm]{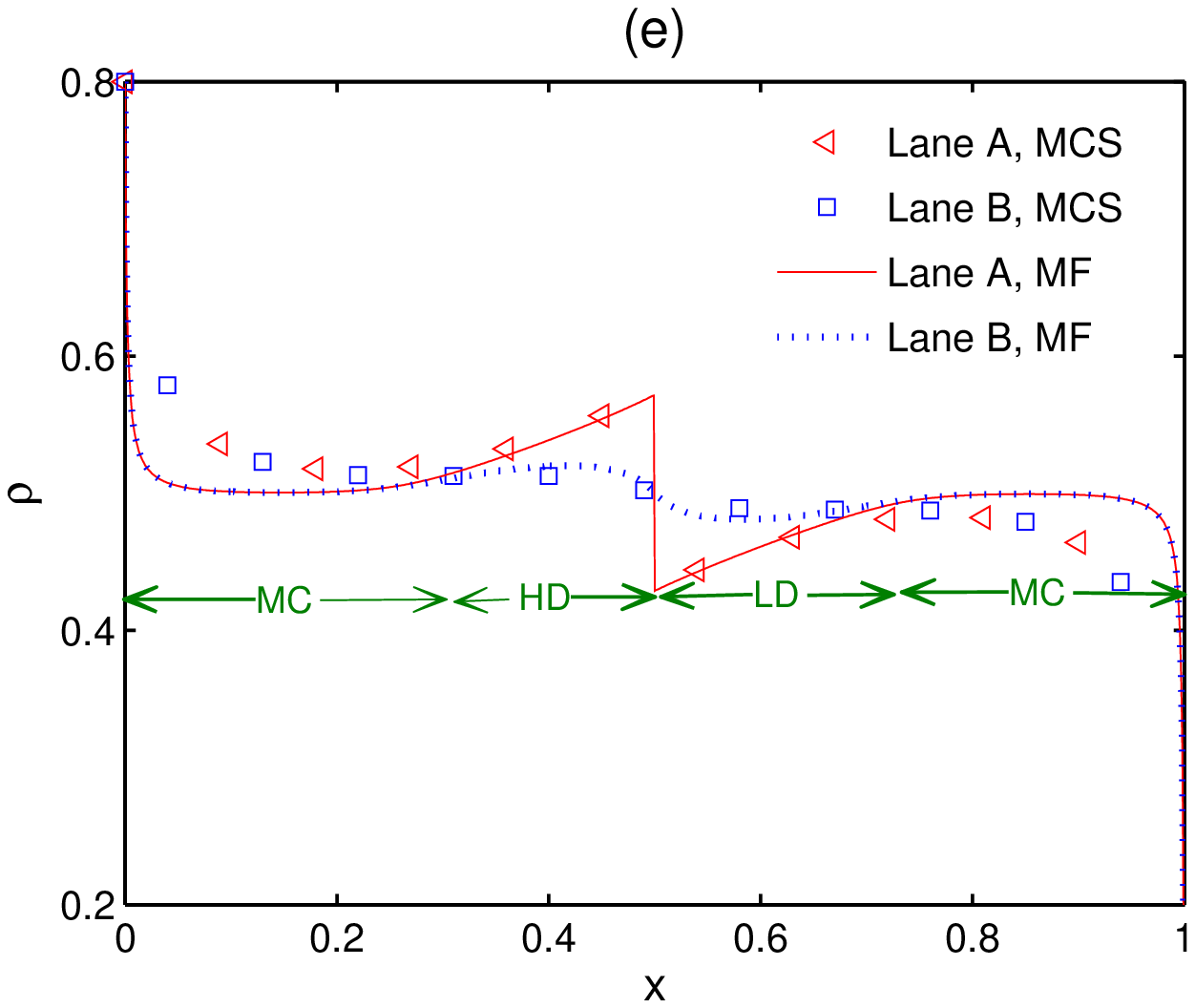}
   \includegraphics[trim=15 00 20 00,clip,height=4.75cm,width=5.25cm]{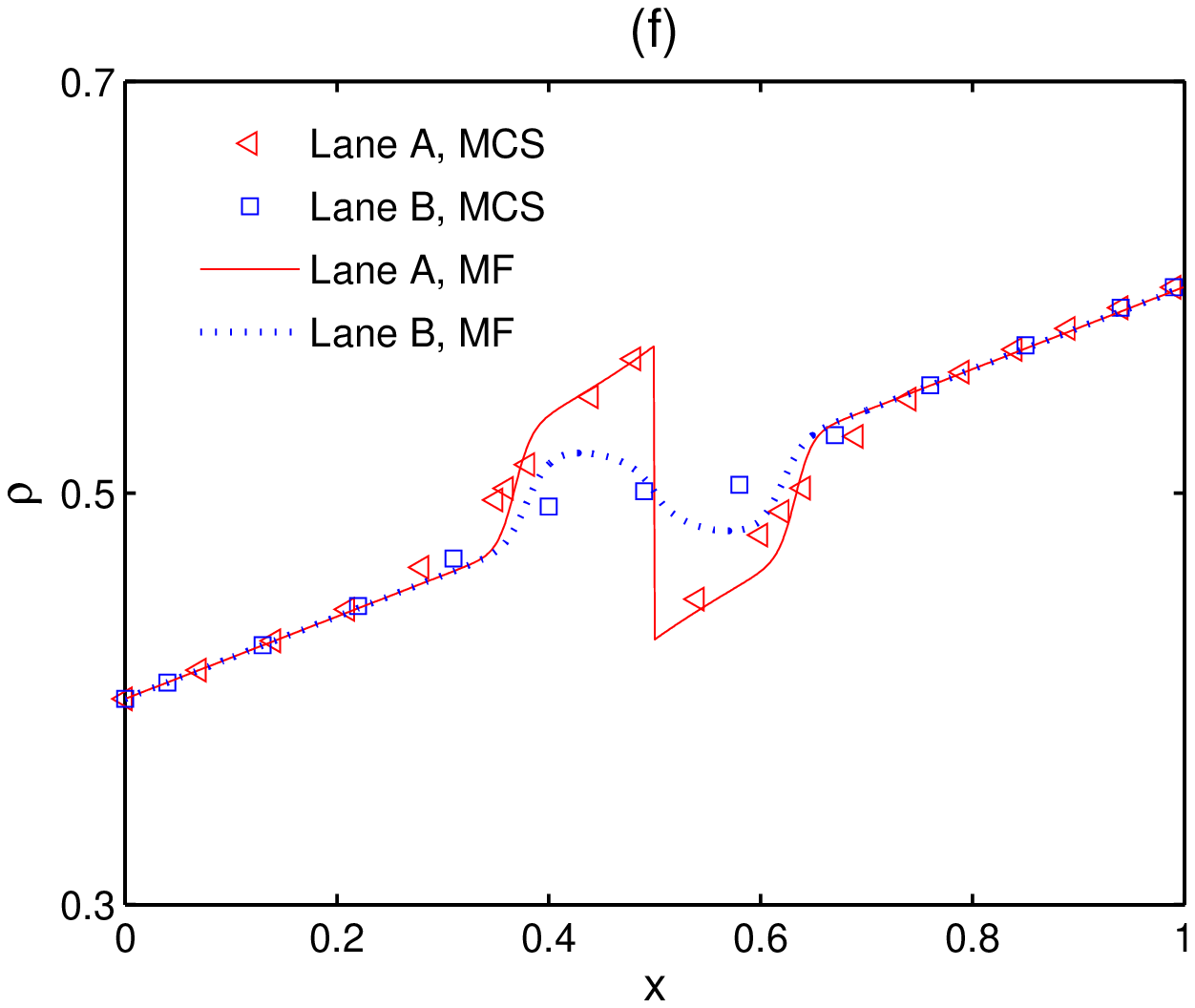}
  \caption{Density profiles for $q=0.75$, $\Omega_d=0.2$, $\Omega=1$ and $L=1000$. (a) (LD-LD,LD/LD) phase for $\alpha=0.2, \beta=0.8$, (b) (LD-S,LD/S) phase for $\alpha=0.1, \beta=0.2$, (c)(HD-HD,HD/HD) for $\alpha=0.5, \beta=0.15$, (d) (S-HD,S/HD) phase for $\alpha=0.2, \beta=0.1$, (e) (MC/HD-LD/MC,MC/HD/LD/MC) phase for $\alpha=0.8, \beta=0.8$ and (f) ($S_b$-S,LD/HD/LD/HD) phase for $\alpha=0.4, \beta=0.4$.}\label{fig:3}
\end{figure}

We list important features of the steady-state dynamics under the current choice of parameters.
\begin{enumerate}
\item \textit{Meissner phase and bottleneck affected region:} In the upper right quadrant ($\alpha \geq 0.5, \beta \geq 0.5$, shaded blue), the bulk densities in each lane becomes independent of entrance as well as exit rates. This effect is similar to the behavior of the Meissner phase found in superconducting materials~\cite{Hirsch2012origin}. Therefore, we also adopt the name Meissner phase (denoted by M) to identify boundary independent behavior of two-channel system. It is important to mention that Meissner phase is actually a part of certain existing phase, denoted by phase $7$ in fig.~\ref{fig:2}(b). Meissner phase is also reported in the literature for a homogeneous single-channel TASEP with LK~\cite{parmeggiani2004totally}. Note that the density profiles get significantly influenced by bottleneck for a specific range of $\alpha, \beta$. The corresponding region in phase-plane is marked with red dots and referred as bottleneck phase throughout this paper. Similar results also exist for single-channel inhomogeneous TASEP with LK, in which bottleneck phase is identified and analyzed in terms of carrying capacity~\cite{pierobon2006bottleneck}. In contrast, for low entrance and exit rates one recovers the TASEP/LK density profiles perturbed by a local spike or a dip, which will be discussed as we proceed in this paper.
\item \textit{Phase transitions and mixed phases:} To gain an intuitive understanding of the steady-state dynamics of the system, it is important to analyze the phase transitions. Across the bulk transition line between (LD-LD,LD)(fig.~\ref{fig:3}(a)) and (LD-S,S) (fig.~\ref{fig:3}(b)) phases, one finds the formation of a shock through deconfinement of the right boundary layer~\cite{gupta2014asymmetric}. It is evident from fig.~\ref{fig:2}(b) that $\beta=\alpha+\Omega_d=\alpha+0.2$ and $\beta=\alpha-\Omega_d=\alpha-0.2$ for $\alpha, \beta \leq 1/2$ are the lines of deconfinement of shock from right boundary layer (LD phase) and left boundary layer (HD phase), respectively. Interestingly, these are also the corresponding lines of deconfinement in analogous single-channel homogeneous system~\cite{mukherji2006bulk}. From here, one can infer that the phase diagram under a weak bottleneck has almost similar topological structure as for the one with no bottleneck ($q=1$)~\cite{parmeggiani2004totally,wang2007effects} though some of the phases differ in their qualitative nature. Actually, a weak bottleneck is incapable of affecting the steady-state dynamics of a symmetrically coupled two-channel TASEP with LK system globally. The only effects can be seen in the vicinity of the bottleneck. This point is evident from the density profiles in fig.~\ref{fig:3}, which will be discussed in detail subsequently.

As one can see from fig.~\ref{fig:3}(b), a shock is present in both the lanes in (LD-S,S) phase. On decreasing $\beta$, the shock in both the lanes moves leftwards and reaches exactly at $x=1/2$ on the line $\alpha=\beta$; $\alpha,\beta \leq 1/2$. This line corresponds to the phase transition between (LD-S,S) and (S-HD,S) phases. A further reduction in magnitude of $\beta$ moves the shock upstream to the bottleneck giving rise to (S-HD,S) phase. The bulk phase transition from (HD-HD,HD)(fig.~\ref{fig:3}(c))to (S-HD,S)(fig.~\ref{fig:3}(d)) can also be understood on the similar lines. Rest of the phase transitions in fig.~\ref{fig:2}(b) are due to gradual changes in the magnitude of densities with respect to entrance and exit rates.

Note that lane B has exactly the same density as that in lane A everywhere, except in the vicinity of the bottleneck (see fig.~\ref{fig:3}). Interestingly, this slight difference in density near the bottleneck is able to produce peculiar kinds of density profiles, giving rise to a number of coexistence (or mixed) phases such as (LD/MC/HD-S,LD/MC/HD/LD/HD) and (MC/HD-LD/MC,MC/HD/LD/MC) etc. An illustration of a mixed phase is shown in fig.~\ref{fig:3}(e).

\item \textit{Formation of upward and downward spikes:} Consider the LD phase in left subsystem of lane A, where the particles have enough space to move along the lattice and the presence of a weak bottleneck does not affect the particle motion significantly. Here, the density profile in lane A incurs a local perturbation in the form of an upward spike as shown in fig.~\ref{fig:3}(a). Due to comparatively slower particle hopping rate at the bottleneck, the particles coming from the left feels a little congestion before entering the bottleneck. Due to this, the density in lane A rises slightly just upstream to the bottleneck, leading to formation of an upward spike in the density profile. Importantly, lane B remains unaffected in this phase. Similarly, one can understand the formation of a downward spike shown in fig.~\ref{fig:3}(b)) in HD phase. This can be attributed to particle-hole symmetry. When a particle comes out of the bottleneck in a highly packed situation, it experiences a sudden increase in its hopping rate. This leads to formation of a lower density region just downstream to the bottleneck. Similar findings have been reported in literature~\cite{pierobon2006bottleneck,wang2008local}.

\item \textit{Transition to bottleneck-induced shock:} Now we discuss an important feature of our system dynamics i.e.~the formation of a bottleneck-induced shock. With LD in lane A, an increase in entrance rate $\alpha$ increases the number of particles in left subsystem of lane A, which ultimately increases height of the existing upward spike. At a particular $\alpha=\alpha_{c}=0.3$, this local density perturbation converts into a shock, which travels in the left subsystem in lane A for further increase in $\alpha$. Since this shock emerges due to bottleneck only, we call this shock as \emph{bottleneck-induced shock} and denote it by $S_b$. The formation of bottleneck-induced shock is also observed for a single-channel inhomogeneous TASEP with LK~\cite{pierobon2006bottleneck}. The presence of bottleneck-induced shock leads to the existence of three new phases ($S_b$-S,LD/HD/LD/HD), ($S_b$-LD/MC/HD,LD/HD/LD/MC) and ($S_b$-LD/MC/HD,LD/HD/LD/
     MC/HD) in the phase-plane as shown in fig.~\ref{fig:2}(b). The existence of mixed phases in lane B indicates the effect of bottleneck in homogeneous lane as well. Interestingly, there exist two shocks in lane A in ($S_b$-S,LD/HD/LD/HD) phase (See fig.~\ref{fig:3}(f)), the left one being bottleneck-induced; whereas the right one is formed due to deconfinement of right boundary layer.
\end{enumerate}
\subsubsection{Effect of bottleneck strength}
We propose a terminology to identify the strength of the bottleneck as shown in table~\ref{tab:1}. We also provide the number of steady-state phases observed under each case, the details of which follow in the text.
\begin{table}[htb]
\caption{Steady-state properties of the system with $\Omega_d=0.2$, $\Omega=1$. $K=3$ is chosen as a specific value for the case $K \ne 1$.}
\label{tab:1}
\centering      
\begin{tabular}{cccc}
\hline\noalign{\smallskip}
\multicolumn{1}{c}{Strength} &
  \multicolumn{1}{c}{Bottleneck}&
   \multicolumn{2}{c}{Number of stationary phases}\\ \cline{3-4}
   \multicolumn{1}{c}{} & \multicolumn{1}{c}{rate} &\multicolumn{1}{c}{$K\ne1$} &\multicolumn{1}{c}{$K=1$}\\ \cline{1-4}
No bottleneck &$q = 1$ & 4 & 7\\
\hline
 Weak  &$0.75 \leq q < 1$ & 2 & 13\\
 \hline
Moderate &$0.4 < q < 0.75$ & 2 & 9\\
\hline
Strong &\makecell{$0.1 < q \leq 0.4$ \\ $0 < q \leq 0.1$}  &\makecell{4  \\ 3 } & 14\\
\hline
Blockage &$q = 0$ & 3 & 14\\
\noalign{\smallskip}\hline
\end{tabular}
\end{table}

\textit{Weak bottleneck:} So far, we have discussed the steady-state dynamics for a weak bottleneck and pointed out important features such as appearance of Meissner phase, bottleneck phase and bottleneck-induced shock etc. Clearly, the number of steady-state phases under a weak bottleneck is more than in homogenous case, which is due to the appearance of mixed phases. Now, we will gradually change the strength of bottleneck from weak to moderate and then to strong and see whether the same trend continues or not?

\begin{figure}[H]
   \includegraphics[trim=10 00 25 00,clip,width=4.5cm,height=4.25cm]{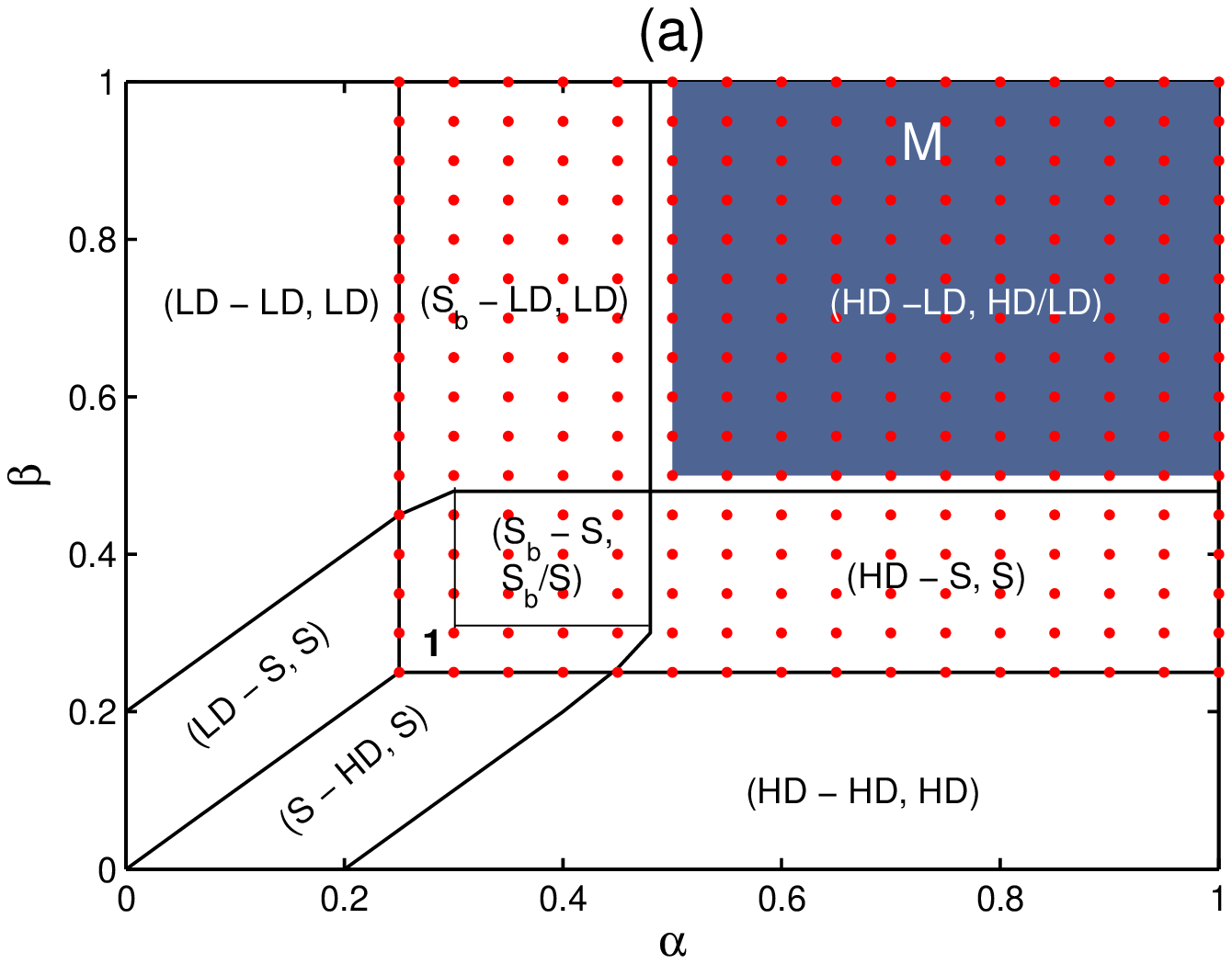}
    \includegraphics[trim=10 00 25 00,clip,width=4.5cm,height=4.25cm]{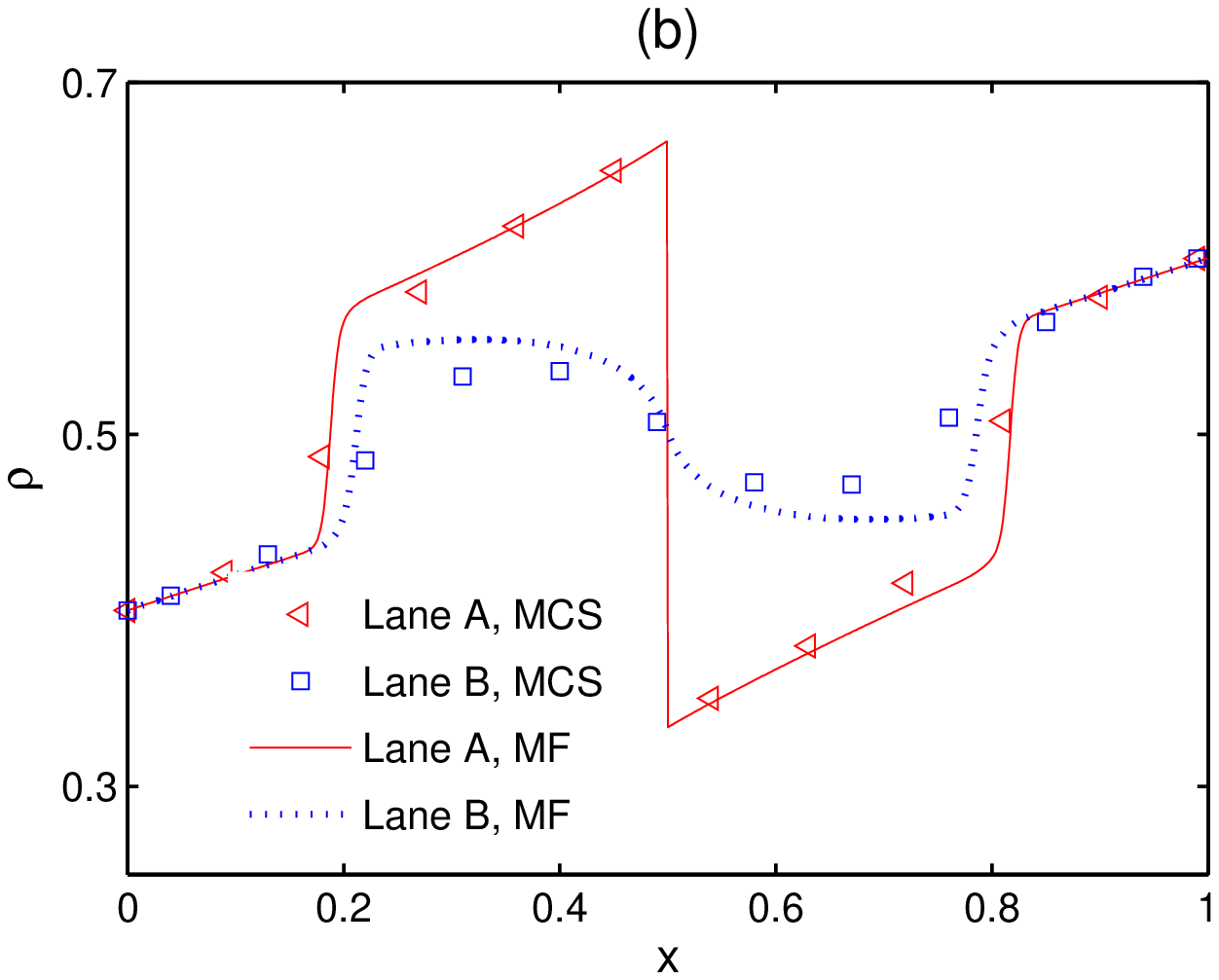}
     \includegraphics[trim=10 00 25 00,clip,width=4.5cm,height=4.25cm]{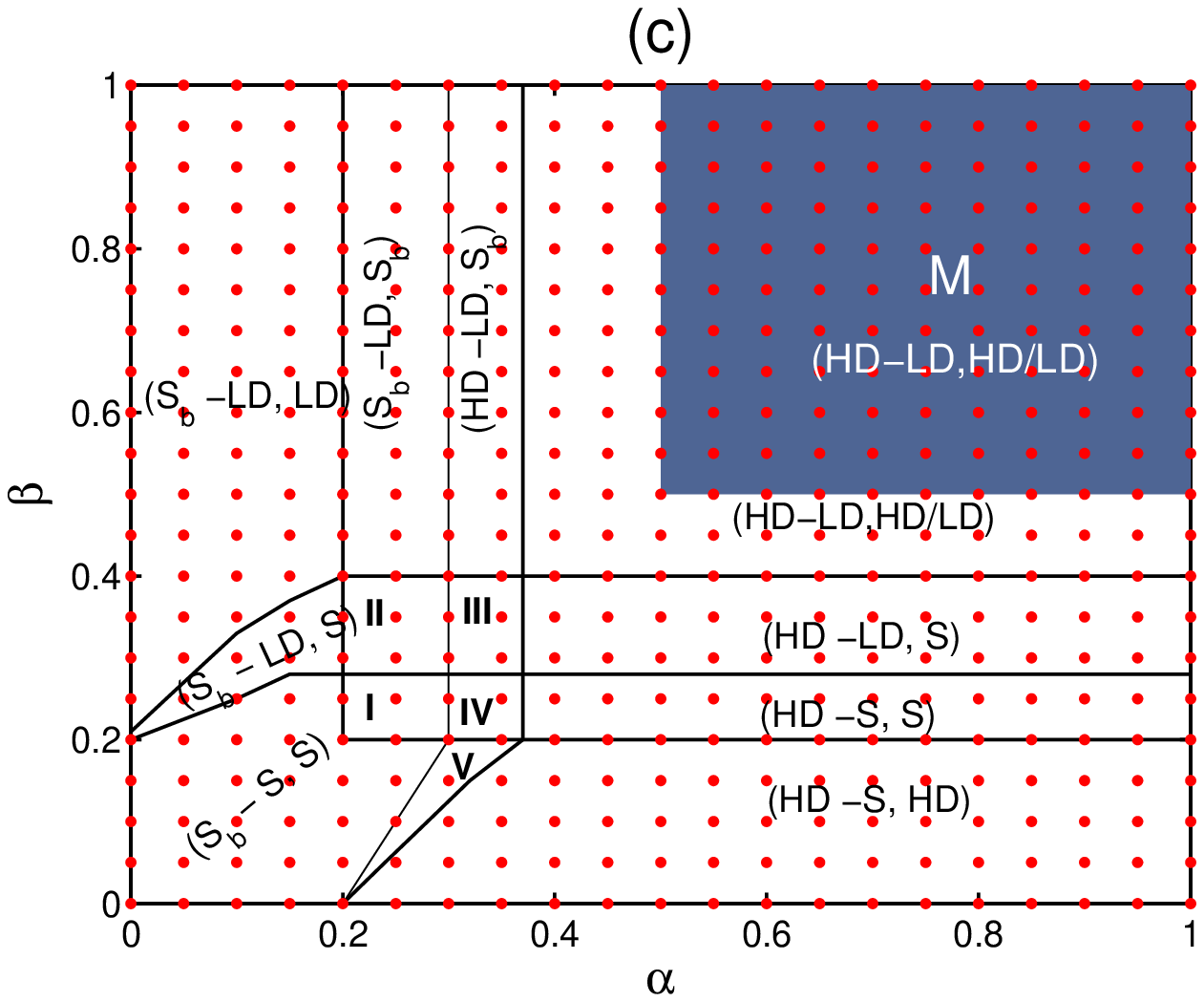}
\caption{Phase diagrams for $\Omega_d=0.2$, $\Omega=1$ and $L=1000$ with (a)$q=0.5$ ,1:($S_b$-S,S), and (b)density profile in ($S_b$-S,$S_b$/S) phase with $\alpha=0.4$, $\beta=0.4$ and $q=0.5$, and (c) phase diagram for $q=0.1$, I: ($S_b$-S,$S_b$/S), II: ($S_b$-LD,$S_b$/S), III: (HD-LD,$S_b$/S) and IV: (HD-S,$S_b$/S). Increase in height of upward spike (shown in magnified inset) and conversion of spike into bottleneck-induced shock with a decrease in $q$}\label{fig:4}
\end{figure}

\textit{Moderate strength:} In the range of moderate strength of bottleneck, we have chosen $q=0.5$ to generate the phase diagram. It is found that for a moderate bottleneck, the phase diagram (See fig.~\ref{fig:4}(a)) is qualitatively as well as quantitatively different from the one with weaker bottleneck (See fig.~\ref{fig:2}(b)).
An important distinguishing feature of the phase diagram is the appearance of bottleneck-induced shock in lane B as well. This can be seen from fig.~\ref{fig:4}(a), which identifies ($S_b$-S,$S_b$/S) as a new phase. While the nature of bottleneck-induced shock in lane A has already been discussed, one needs to investigate the occurrence of two shocks in lane B. Clearly one located in the right is due to the deconfinement of right boundary layer; while other is because of the indirect impact of bottleneck in lane A, whose origin can be understood as follows. Actually, the presence of a bottleneck-induced shock in lane A produces a local high-density region there, making it difficult to accommodate incoming particles from lane B. As a result, the density in the left subsystem of lane B rises and converts into a shock with an increase in entrance rate $\alpha$. The shock produced in lane B, due to the effect of bottleneck in lane A, is also called as bottleneck-induced shock.

On moving from weak to moderate bottleneck, the phase diagram gets simplified resulting into a reduction in number of phases. The coexistence phases (MC/HD-LD/MC,MC/HD/LD/MC), (MC/HD-S,MC/HD/LD/HD) and ($S_b$-LD/MC,LD/HD/LD/MC), observed in fig.~\ref{fig:2}(b), are replaced by (HD-LD,HD/LD), (HD-S,S) and ($S_b$-LD,S), respectively. This shows that MC phase, either as a whole or as a part of density profile, does not exist for moderate bottleneck, contrary to the weak bottleneck case. This result is analogous to the one found for single-channel TASEP with LK with a single bottleneck~\cite{pierobon2006bottleneck}. Further, the bottleneck phase (dotted region) enlarges, as expected for a stronger bottleneck; whereas the phases (LD-LD,LD),(LD-S,S), (S-HD,S) and (HD-HD,HD) shrink. Meissner phase remains unaffected with increase in bottleneck strength.

\textit{Strong bottleneck:} Fig.~\ref{fig:4}(c) shows the phase diagram for the present system with $q=0.1$, a strong bottleneck. It can be seen that topology of the phase diagram is significantly changed in comparison to the one with moderate bottleneck. Here, the bottleneck effect is sufficiently strong to increase the density difference between two lanes up to a level that we have a complex phase diagram structure as shown in fig.~\ref{fig:4}(c). One can see that LD phase does not exist in the left subsystem to the bottleneck in lane A. This is physically justified as the extremely slow hopping rate at bottleneck obstructs the incoming particles, leading to formation of either a shock or a HD region in left subsystem. Similar argument and particle-hole symmetry accounts for the non-existence of HD phase downstream to the bottleneck in lane A. It further accounts for the enlargement of (HD-LD,HD/LD) phase with a decrease in $q$. Note that the bottleneck phase (dotted region) further enlarges to maximum extent and covers the whole $\alpha-\beta$ plane.

We have also checked the results for the case of complete blockage at the bottleneck i.e. $q=0$. The phase diagram for $q=0$ remains essentially similar to the one for $q=0.1$. The only change is shifting of the phase boundaries, which leads to expansion of (HD-LD, HD/LD) phase (following the similar trend as explained above) and shrinkage of rest of the phases.
\begin{figure}
\centering
\includegraphics[trim=10 00 25 00,clip,width=6.5cm,height=5.25cm]{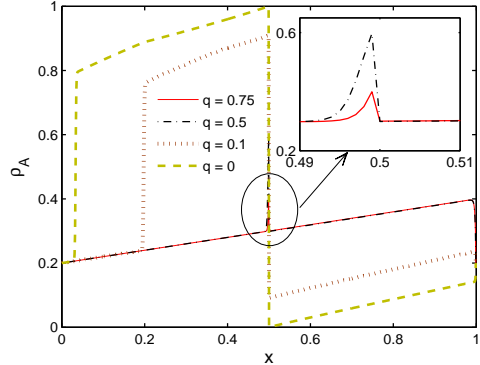}
\caption{Increase in height of upward spike (shown in magnified inset) and conversion of spike into bottleneck-induced shock with a decrease in $q$.}\label{fig:5}
\end{figure}

Fig.~\ref{fig:5} shows the transition of upward spike into bottleneck-induced shock with respect to an increase in bottleneck strength. Firstly the height of upward spike in lane A at $x=1/2$ increases in moderate bottleneck case(clearly seen in the inset), then the spike converts into shock for strong bottleneck. A further decrease in value of $q$ pulls the shock leftwards.
\subsubsection{Effect of coupling strength}
So far, we have investigated the system dynamics under different bottleneck strengths, but for fixed lane-changing rate at $\Omega=1$. Since coupling strength has always been an important parameter in two-channel systems, it is necessary to examine its role for the proposed model. In literature, it is well explored~\cite{dhiman2014effect,gupta2013coupling,pronina2004two,wang2007effects} that higher orders of lane-changing rate does not affect the dynamics of a symmetrically coupled homogeneous two-channel TASEP with/without LK because the system decouples into two independent single-channel homogeneous systems. But, in our model, this is not the case as our system is inhomogeneous. So, we thoroughly investigate the effect of coupling strength on the dynamics of the proposed inhomogeneous model under different strengths of bottleneck systematically. Our aim is to discuss the important topological changes in the phase diagram induced by higher order of lane-changing rate (For definition of order, please see~\cite{dhiman2014effect}).

\begin{figure}[htb]
\centering
 \includegraphics[width=6.75cm,height=5.8cm]{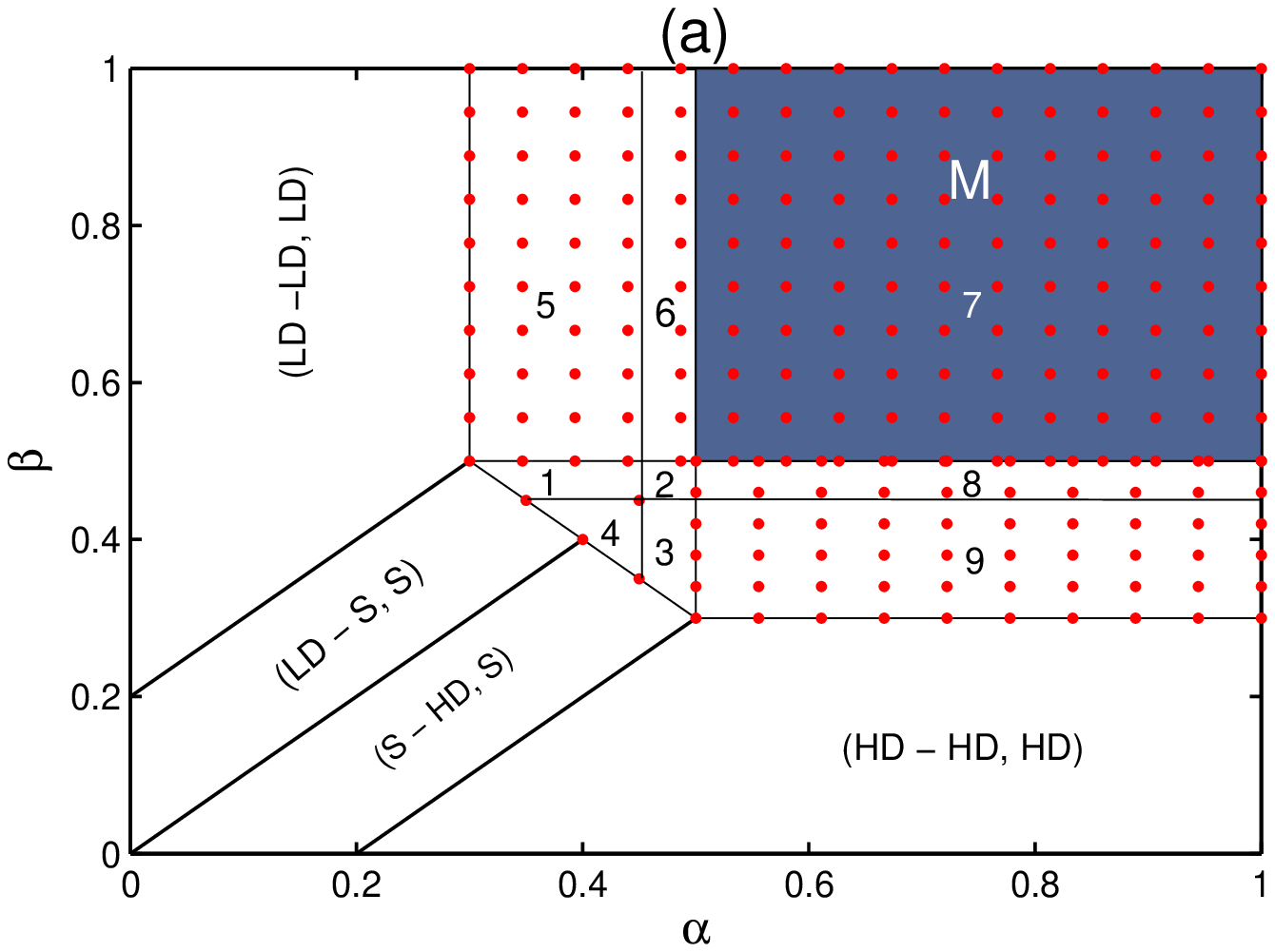}
    \includegraphics[width=6.75cm,height=5.8cm]{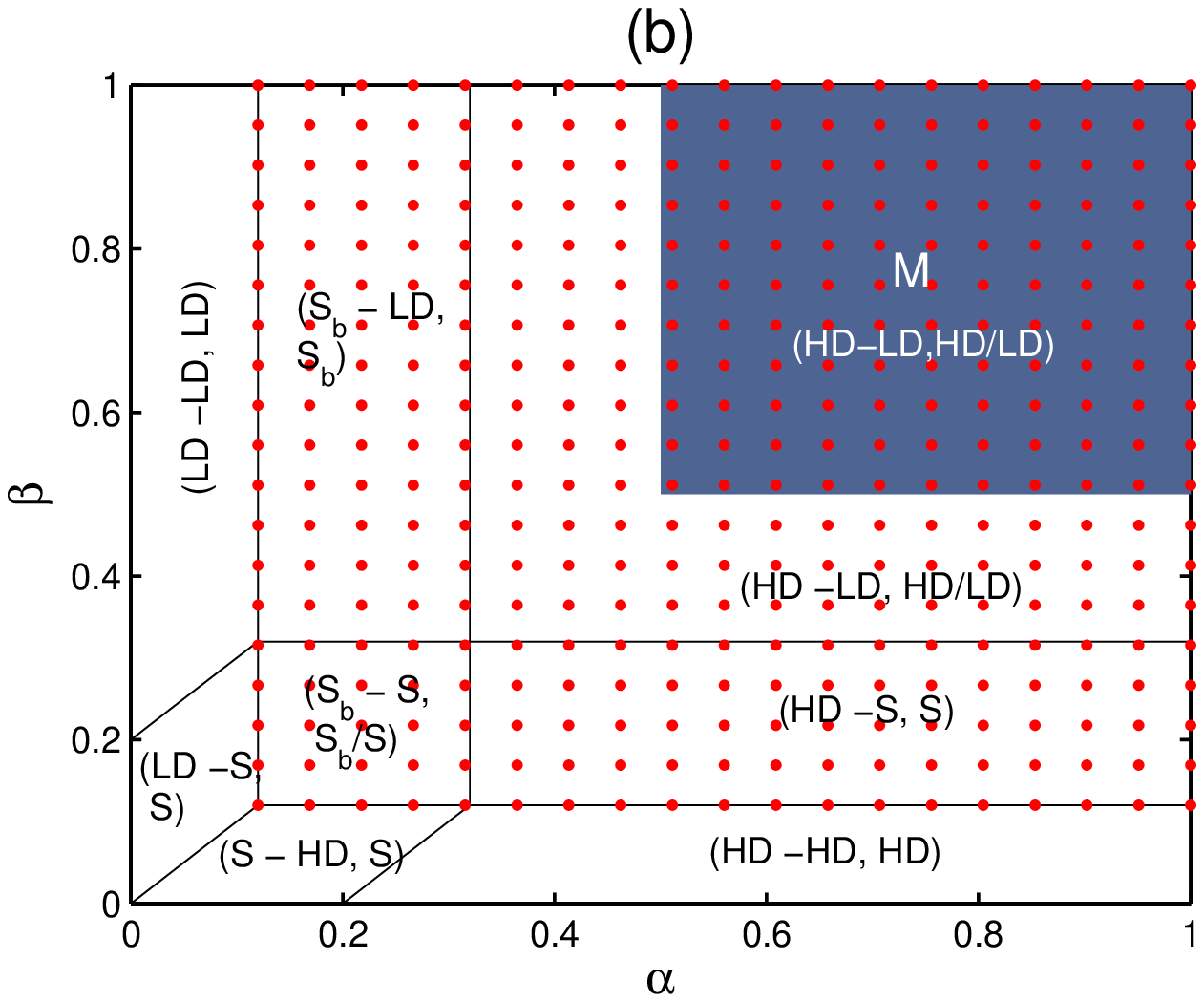}
  \caption{Phase diagram for $\Omega_d=0.2$, and $L=1000$ with (a) $q=0.75$, $\Omega=100$, Notations:- 1: ($S_b$-LD/MC/HD,$S_b$/LD/MC/HD), 2: (LD/HD-LD/MC/HD,LD/HD/LD/MC/HD), 3: (LD/HD-HD,LD/HD) and 4: ($S_b$-HD,$S_b$/HD).(b) $q=0.1$ with $\Omega=10$. }\label{fig:6}
\end{figure}

For a weak bottleneck, the effect of lane-changing rate on the phase diagram is insignificant till $O(\Omega) \leq 100 O(\Omega_d)$. Only little quantitative differences are seen on density profiles in both the lanes. When $O(\Omega) > 100 O(\Omega_d)$, we notice significant changes in the density profiles of both the lanes leading to some qualitative changes in the nature of certain phases. Fig.~\ref{fig:6}(a) shows the phase diagram for $q=0.75$ with $\Omega=100$. For higher orders of $\Omega$, we have found that the number of stationary phases remain same. Interestingly, some of the existing phases get transformed into new phases without a noticeable change in the phase boundaries. For example, the phases (LD/MC/HD-LD/MC,LD/MC/HD/LD/MC), (MC/HD-LD/MC,MC/HD/LD/MC) and

($S_b$-S,LD/HD/LD/HD) change to (LD/HD-MC,LD/HD/MC), (MC/HD-MC,
MC/HD/MC), ($S_b$-HD,$S_b$/HD), respectively on moving from $\Omega=1$ to $\Omega=100$.

The effect of increasing the order of $\Omega$ on a moderate bottleneck is similar to the one discussed for weak bottleneck. Therefore, we skip the phase diagrams for this case at higher orders of $\Omega$.

Moving our attention towards strong bottleneck, we analyze the phase diagram for $q=0.1$. No structural changes in the phase diagram are seen till $O(\Omega) < 10 O(\Omega_d)$, except minor translation of various phase boundaries. As an example, the ($S_b$-S,S) phase expands; while (HD-LD,HD/LD) phase shrinks. An important phenomenon occurs at $O(\Omega) = 10 O(\Omega_d)$, where the density difference between both the lanes, in the vicinity of the bottleneck, decreases. As a result, the complexity of the phase diagram reduces as shown in fig.~\ref{fig:6}(b). Importantly, four new phases (LD-LD,LD), (LD-S,S), (S-HD,S) and (HD-HD,HD) appear. The effect of bottleneck, despite its higher strength at $q=0.1$, reduces because the bottleneck phase covers comparatively smaller region in $\alpha-\beta$ plane. Further increase in lane-changing rate does not produce any subsequent changes in the phase diagram for $q=0.1$. This happens due to the weakening of bottleneck effect by an increase in lane-changing rate, which is discussed as follows.

\begin{figure}[htb]
\centering
 \includegraphics[width=6.75cm,height=5.8cm]{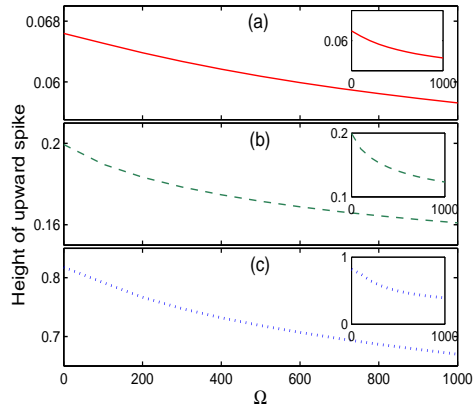}
 \caption{Variation in height of upward spike (insets show for downward spike) at bottleneck in lane A for (a) $q=0.75$, (b) $q=0.5$ and (c) $q=0.1$. Here, $\alpha=0.1$, $\beta=0.6$ for upward spike in LD phase and $\alpha=0.8$, $\beta=0.1$ for downward spike in HD phase.}\label{fig:7}
\end{figure}

\textit{Weakening of bottleneck effect}
It can be seen from figs.~\ref{fig:2},~\ref{fig:4} and ~\ref{fig:6} that the bottleneck affected region (dotted region) shrinks at higher orders of lane-changing rate in comparison to the one for lower orders of $\Omega$ under all values of $q$. Specifically, the weakening of bottleneck effect occurs for lower values of $\alpha$ and $\beta$ with an increase in lane-changing rate $\Omega$. This is further justified from fig.~\ref{fig:7}, which shows a monotonic reduction in peak's height of upward spike at bottleneck with respect to an increase in $\Omega$. An obvious explanation of this effect is that the particles in lane A, feeling hindered due to the bottleneck, would shift to other lane more frequently. As a result, the congestion in lane A gets slightly reduced as one increase the lane-changing rate. Similar kind of effect is also reported in literature~\cite{wang2007effects}. The insets in fig.~\ref{fig:7} shows the similar effect for downward spike as well.
%

\subsection{$K\neq1$}
Since the methodology of the proposed hybrid mean-field approach is general and applicable for any value of binding constant $K$, we follow same lines to discuss the case where attachment and detachment rates differ, $\Omega_a \ne \Omega_d$, i.e. $K \ne 1$.
\begin{figure}[htb]
\includegraphics[trim=10 00 25 00,clip,width=4.5cm,height=4.25cm]{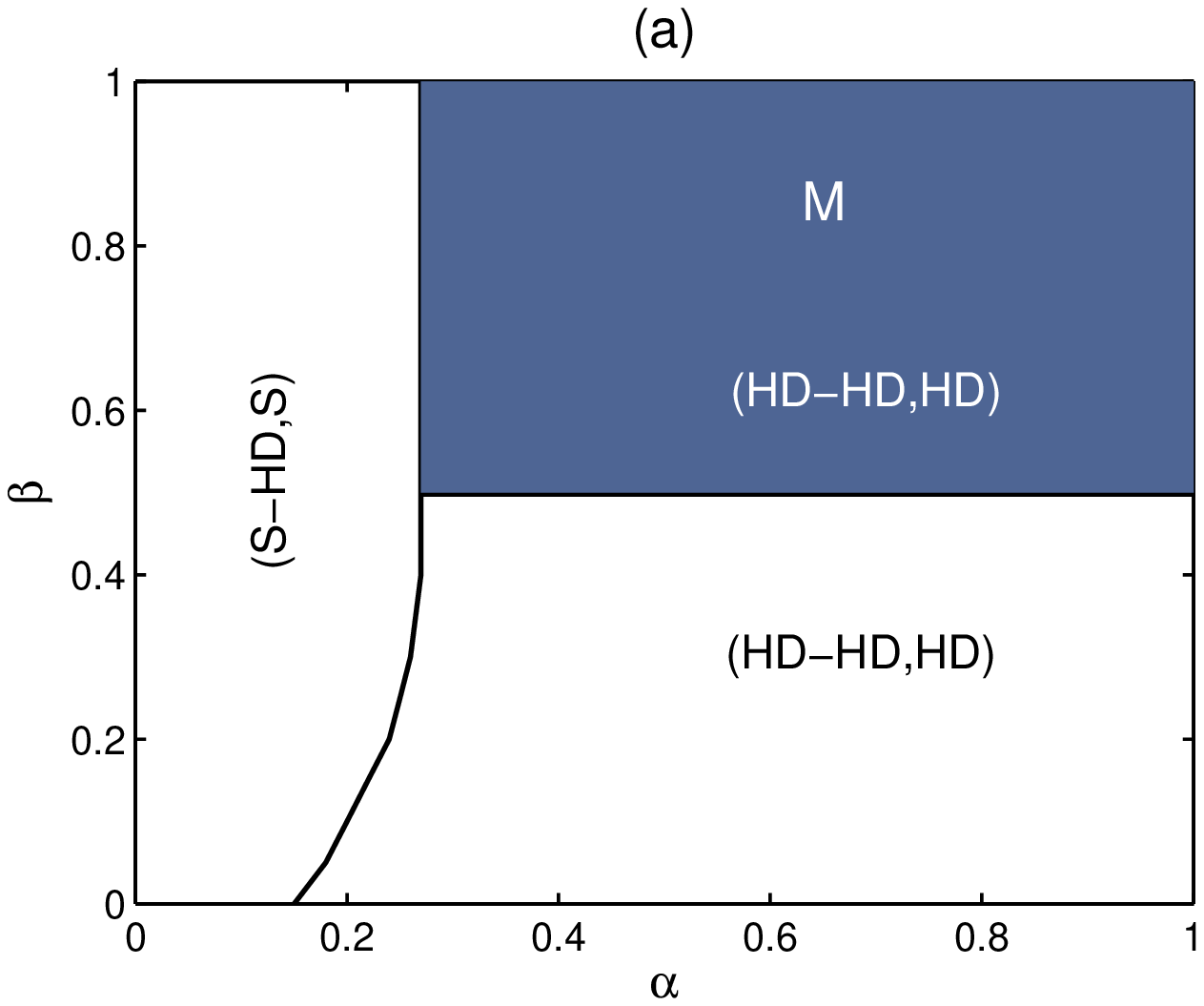}
\includegraphics[trim=10 00 25 00,clip,width=4.5cm,height=4.25cm]{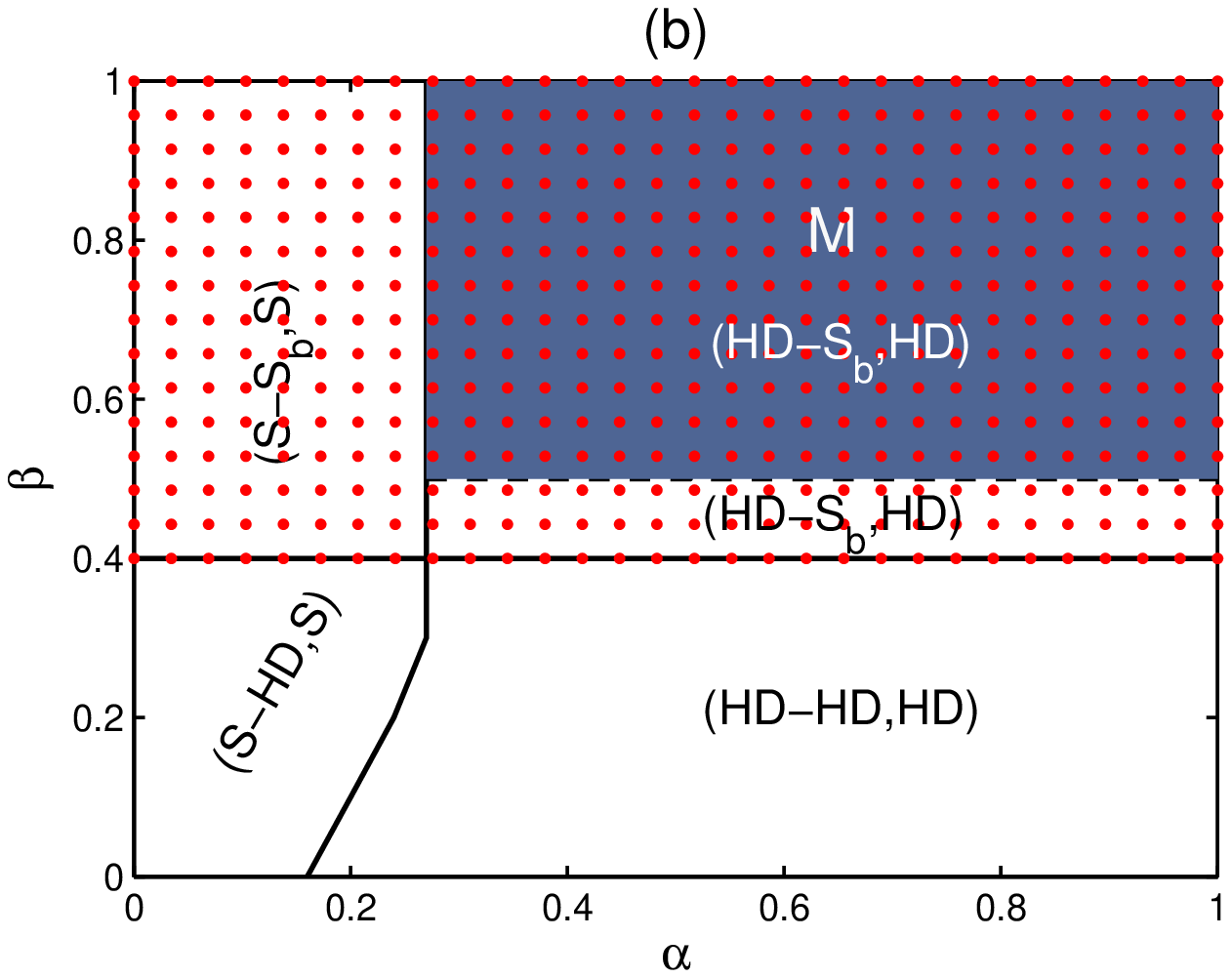}
\includegraphics[trim=10 00 25 00,clip,width=4.5cm,height=4.25cm]{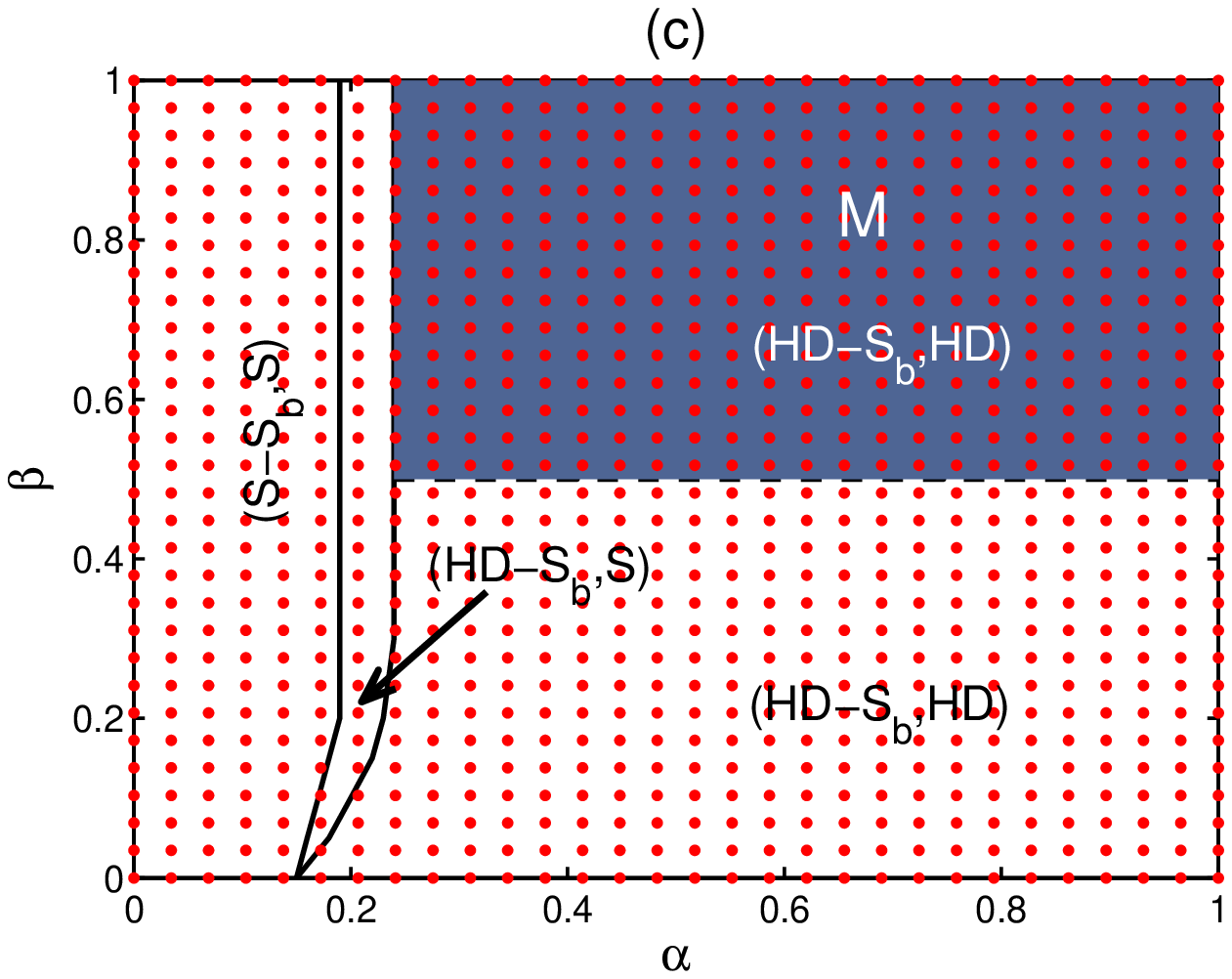}
\caption{The phase diagrams for $K=3$, $\Omega_d=0.2$ and $\Omega=0.5$ with different $q$: (a) $q=0.75$, (b) $q=0.4$, and (c) $q=0.1$. Notations, M: Meissner phase (shaded blue) and BP: Bottleneck phase (shaded with red dots)}\label{fig:8}
\end{figure}

Fig.~\ref{fig:8} shows the steady-state phase diagrams with $K=3$. Comparing fig.~\ref{fig:8} with figs.~\ref{fig:2} and ~\ref{fig:4}, we notice that the complexity of the phase diagrams reduces tremendously for $K=3$ under all values of $q$. The number of steady-state phases is shown in table~\ref{tab:1}. Note that due to the increased attachment rate in comparison to detachment rate, the density in both the lanes increases, leading to a complete absence of LD phase in the system.

For weaker bottleneck, the phase diagram is much simple consisting of only two stationary phases (S-HD,S) and (HD-HD,HD) as shown in fig.~\ref{fig:8}(a). Quite surprisingly, no density profile in $\alpha-\beta$ plane is significantly affected by bottleneck, excluding the presence of a downward spike at $x=1/2$. This is in sharp contrast to the parallel case with $K=1$ (See fig.~\ref{fig:2}(b)). Consequently, one does not get a bottleneck-induced shock here for any values of $\alpha$ and $\beta$.

Further decrease in the value of $q$ does not change the structure of phase diagram but increases the height of downward spike in HD phase. This trend continues till one reaches $q_{c,2}\approx 0.4$, at which the perturbation of downward spike enters the right subsystem in lane A in the form of a shock, called as bottleneck-induced shock. Its emergence leads to formation of new steady-state phases such as ($S-S_b,S$) and (HD-$S_b$,HD) as shown in fig.~\ref{fig:8}(b). A part of phase-plane (shaded with red dots) is strongly influenced by the bottleneck-induced shock called as bottleneck phase, which was absent for $q > 0.4$ (See fig.~\ref{fig:8}(a)). On the other hand, the value of $q$ does not influence the Meissner phase.

A further decrease in value of $q$ below $q_{c,2}$ leads to significant changes in the steady-state dynamics. The emergence of a new phase (HD-$S_b$,S) is seen at another critical value $q_{c,3}=0.1$ as shown in fig.~\ref{fig:8}(c). Additionally two of the earlier existing phases (S-HD,S) and (HD-HD,HD) disappear completely; while the bottleneck phase covers the whole $\alpha-\beta$ plane reflecting the profound effect of strong bottleneck.
\subsubsection{Effect of coupling strength}
\begin{figure}[htb]
\centering
\includegraphics[trim=10 00 25 00,clip,width=5.25cm,height=4.75cm]{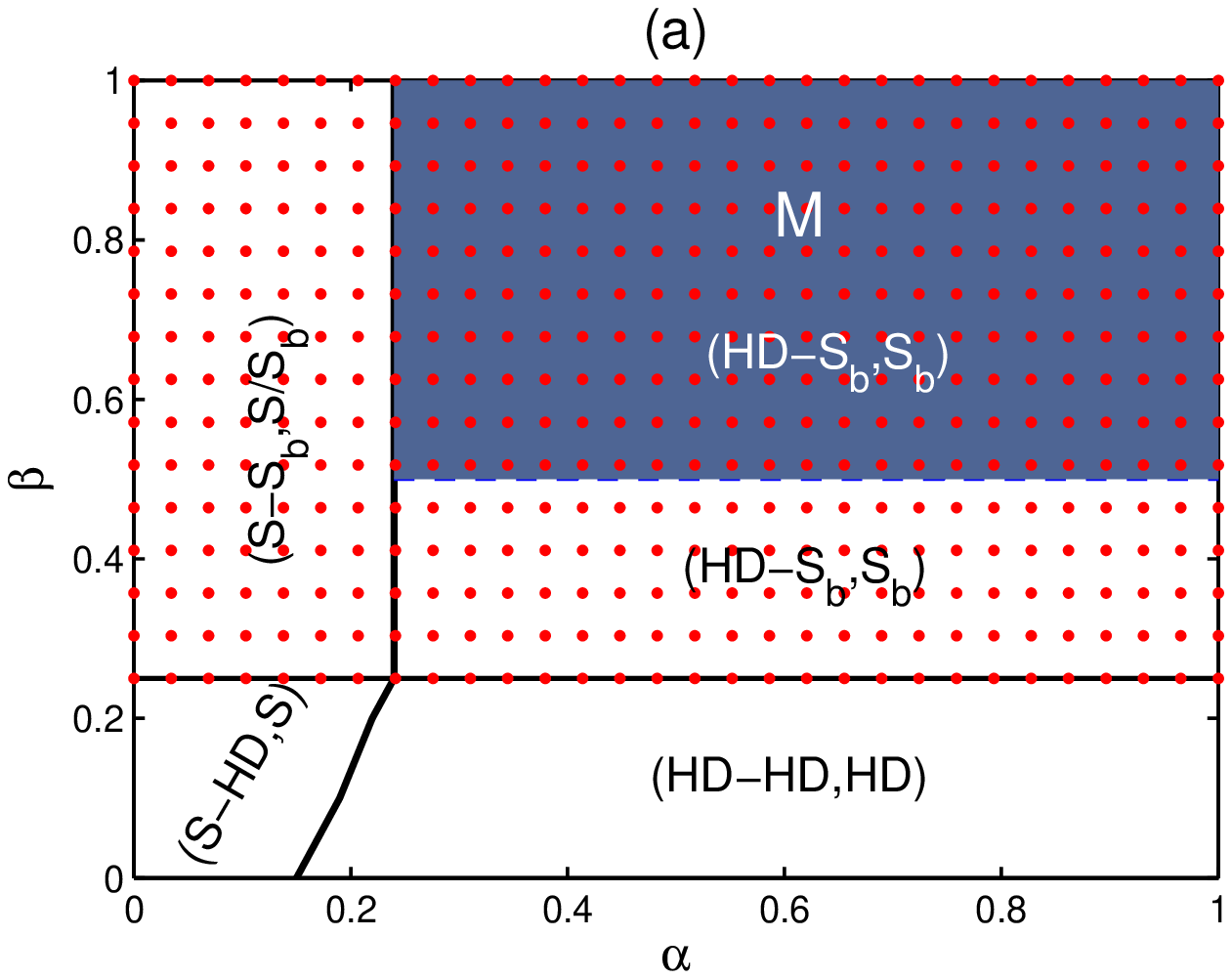}
\includegraphics[trim=10 00 25 00,clip,width=5.25cm,height=4.75cm]{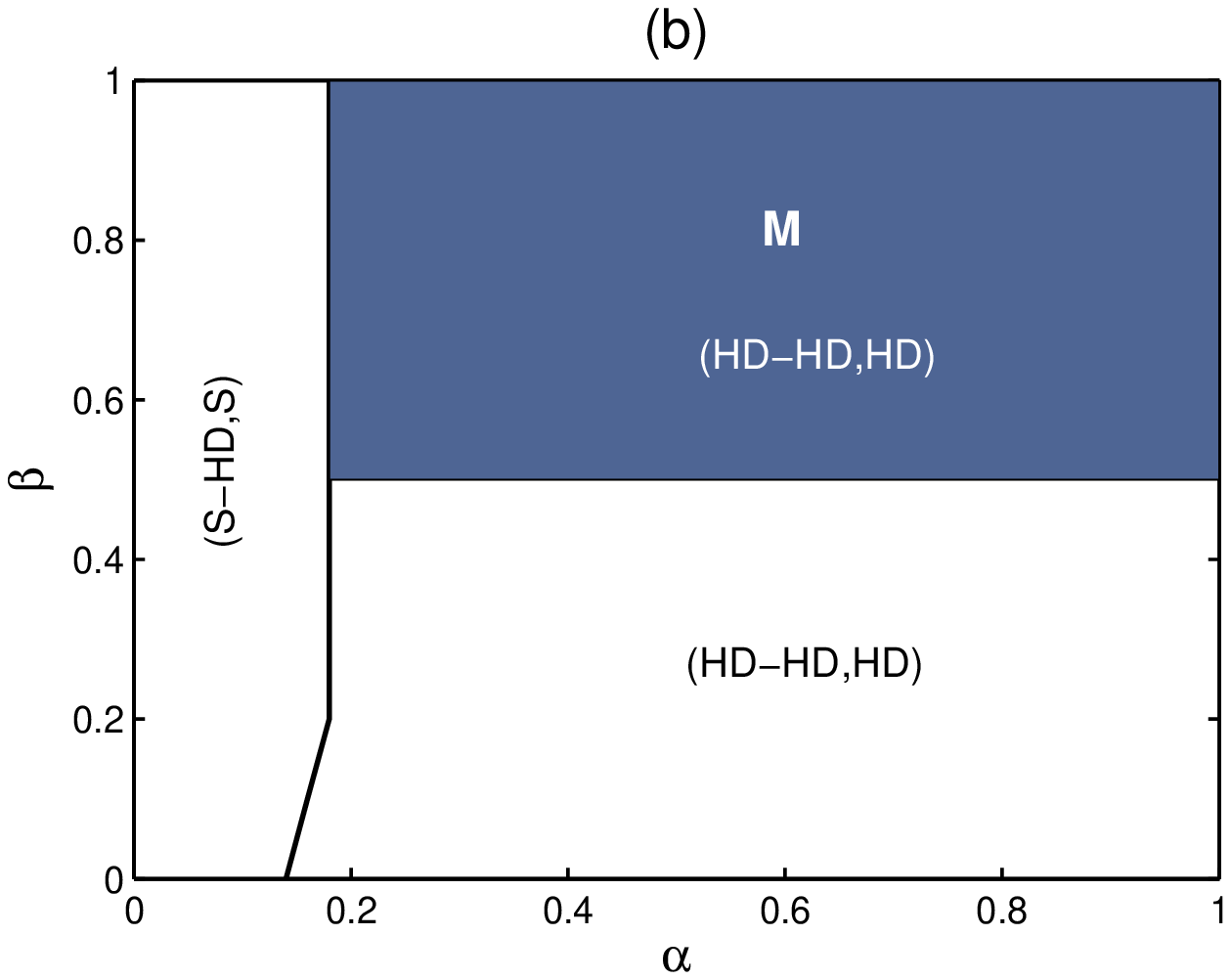}
\caption{Phase diagrams for (a) $q=0.1$, $K=3$, $\Omega=100$ and (b)$q=0.01$, $K=4$, $\Omega=1000$.}\label{fig:9}
\end{figure}

We have also investigated the effect of lane-changing rates on the dynamics of the proposed system under case $K\ne1$. For $O(\Omega) \leq 10 O(\Omega_d)$, no significant changes are seen except shifting of phase boundaries for any value of $q$. We focus on the situation when $O(\Omega) \geq 100 O(\Omega_d)$. It is found that the steady-state phase diagram for weak bottlenecks remains unchanged and has a structure similar to the one seen in fig.~\ref{fig:8}(a). But the height of downward spike at $x=1/2$ decreases with respect to increase in lane-changing rate, analogously to the parallel case $K=1$. One can infer from here that an increase in lane-changing rate weakens the bottleneck effect with $K \ne 1$ as well. The extent of weakening effect is up to a level that we do not get a bottleneck-induced shock till $q > 0.15$, which means the global effects of the disturbance caused by bottleneck can be captured only for a very strong bottleneck. As an illustration, we provide the phase diagram for $q=0.1$ in fig.~\ref{fig:9}(a), which consists of four new stationary phases. Moreover, the bottleneck phase shrinks in its size, which once again indicates the weakening of the bottleneck's effect with respect to an increases in $\Omega$.
\subsection{Non-existence of bottleneck-induced shock}
We have observed that an increase in lane-changing rate weakens the bottleneck effect and as a result the bottleneck affected region in $\alpha-\beta$ plane shrinks with respect to an increase in $\Omega$. Comparing the two cases $K=1$ and $K\ne1$, one can infer that the effect of bottleneck is further reduced when $K \ne 1$. The above two statements generate interest to know whether there exist some values of $K$ and/or $\Omega$, at which we do not get a bottleneck-induced shock even for a strong bottleneck. A thorough analysis of our study has revealed that for $K=4$ and $\Omega=1000$ with a very strong bottleneck ($q=0.01$, almost a blockage), we get a steady-state situation free of any bottleneck-induced shock. The resulting steady-state phase diagram is shown in fig.~\ref{fig:9}(b). Although we do have a downward spike in the density profiles, yet the non-existence of a bottleneck-induced shock is an important outcome. Such dynamics further opens up new challenges to dig deeply into the shock dynamics of the proposed model, which will be undertaken in our future study.
\section{Conclusion}
\label{sec:7}
In this paper, we attempted to provide a clear picture of the role played by an inhomogeneity in a two-channel TASEP with LK under a symmetric coupling environment. The location of bottleneck is fixed at middle site in lane A, while lane B is kept homogeneous. Theoretically, we have introduced a new \emph{hybrid} approach and also employed singular perturbation technique to get the steady-state density profiles and phase diagrams. We have adopted random-sequential update rules to perform Monte Carlo simulations and shown a good agreement between theoretical and MCS.

The study is segmented into two subcases: $K=1$ and $K\ne1$. Based on the analysis, one finds very interesting topologies of the phase diagrams under different bottleneck rates. It is concluded that the steady-state phase diagrams for different bottleneck strengths under $K=1$ are comparatively richer in their composition as compared to those under $K\ne1$. For lower strengths of bottleneck, the phase diagram comes out to be quite similar to the corresponding homogeneous system and effects on the density profiles are only local in the vicinity of the bottleneck. The origin of upward and downward spike at bottleneck in LD and HD phases, respectively is explained. We also shed light on the emergence of bottleneck-induced shock in the system, which accounts for the global effects of bottleneck. The effect of lane-changing rate on the dynamics is also examined and found that an increase in coupling strength weakens the bottleneck effect for any value of $K$. This effect is supported by the reduction in height of upward and downward spikes with respect to increase in lane-changing rate. Another important observation is that unequal attachment-detachment rates are also responsible for weakening of bottleneck effect.

Here, we focused on a localized bottleneck, fixed at the middle of the lattice far away from
the boundaries. The approach developed in this paper might be easily extended to analyze more general
inhomogeneous systems such as those with more number of bottlenecks and also to study the influence
of the location of the bottleneck in the multi-channel TASEPs. From the biological significance, one
can say that the present study might help in optimizing the protein translation rate in multi-channel
inhomogeneous systems.

\section*{Acknowledgements}
A.K. Gupta gratefully acknowledges the financial support from the Department of Science
and Technology (DST), Government of India. I. Dhiman gratefully thanks CSIR, India for
the Ph.D. fellowship.



%
%
%
\end{document}